\documentclass[twocolumn,secnumarabic,amssymb, nobibnotes, aps, pra, superscriptaddress]{revtex4-2}
\pdfoutput=1
\usepackage{CJK}
\usepackage{graphicx}% Include figure files
\usepackage{dcolumn}% Align table columns on decimal point
\usepackage{bm}% bold math

\usepackage[utf8]{inputenc}
\usepackage{color}
\usepackage[T1]{fontenc}
\usepackage{booktabs, array, mathptmx, float, tabularx, booktabs, lipsum, amsmath,multirow}
\usepackage{siunitx, xcolor}
\graphicspath{{figs/}{figsgaoerb/}} % 读取图片从figs/文件夹中读取，而不是当前文件夹
\usepackage[colorlinks,linkcolor=blue,anchorcolor=blue,citecolor=blue,urlcolor=blue]{hyperref}

\begin{document}

% Use the \preprint command to place your local institutional report
% number in the upper righthand corner of the title page in preprint mode.
% Multiple \preprint commands are allowed.
% Use the 'preprintnumbers' class option to override journal defaults
% to display numbers if necessary
%\preprint{}

%Title of paper
\title{Characterizing the spatial potential of a surface electrode ion trap}

% repeat the \author .. \affiliation  etc. as needed
% \email, \thanks, \homepage, \altaffiliation all apply to the current
% author. Explanatory text should go in the []'s, actual e-mail
% address or url should go in the {}'s for \email and \homepage.
% Please use the appropriate macro foreach each type of information

% \affiliation command applies to all authors since the last
% \affiliation command. The \affiliation command should follow the
% other information
% \affiliation can be followed by \email, \homepage, \thanks as well.
\author{Qingqing Qin}
\thanks{These authors contributed equally to this work.}
\author{Ting Chen}
\thanks{These authors contributed equally to this work.}
\affiliation{College of Science, National University of Defense Technology, Changsha 410073, P. R. China}
\affiliation{ Hunan Key Laboratory of Mechanism and Technology of Quantum Information, Changsha 410073, Hunan, P. R. China}

\author{Xinfang Zhang}
\affiliation{  Institute for Quantum Information \& State Key Laboratory of High Performance Computing, College of Computer Science, National University of Defense Technology, Changsha 410073, China}

\author{Baoquan Ou}
\affiliation{College of Science, National University of Defense Technology, Changsha 410073, P. R. China}
\affiliation{ Hunan Key Laboratory of Mechanism and Technology of Quantum Information, Changsha 410073, Hunan, P. R. China}

\author{Jie Zhang}
\author{Chunwang Wu}
\author{Yi Xie}
\email{xieyi2015@nudt.edu.cn}
\author{Wei Wu}
\email{weiwu@nudt.edu.cn}
\author{Pingxing Chen}
\affiliation{College of Science, National University of Defense Technology, Changsha 410073, P. R. China}
\affiliation{ Hunan Key Laboratory of Mechanism and Technology of Quantum Information, Changsha 410073, Hunan, P. R. China}

%Collaboration name if desired (requires use of superscriptaddress
%option in \documentclass). \noaffiliation is required (may also be
%used with the \author command).
%\collaboration can be followed by \email, \homepage, \thanks as well.
%\collaboration{}
%\noaffiliation

\date{\today}

\begin{abstract}
	
The accurate characterization of the spatial potential generated by a planar electrode in a surface-type Paul trap is of great interest. To achieve this, we employ a simple yet highly precise parametric expression to describe the spatial field of a rectangular-shaped electrode. Based on this, an optimization method is introduced to precisely characterize the axial electric field intensity created by the powered electrode and the stray field. In contrast to existing methods, various types of experimental data, such as the equilibrium position of ions in a linear string, equilibrium positions of single trapped ions and trap frequencies, are utilized for potential estimation in order to mitigate systematic errors. This approach offers significant flexibility in voltage settings for data collection, making it particularly well-suited for surface electrode traps where ion probe trapping height may vary with casual voltage settings. In our demonstration, we successfully minimized the discrepancy between experimental observations and model predictions to an impressive extent. The relative errors of secular frequencies were suppressed within ±0.5$\%$, and the positional error of ions was limited to less than 1.2  $\mu$m, all surpassing those achieved by existing methodologies.

\end{abstract}

% insert suggested keywords - APS authors don't need to do this
%\keywords{}

%\maketitle must follow title, authors, abstract, and keywords
\maketitle

% body of paper here - Use proper section commands
% References should be done using the \cite, \ref, and \label commands
\section{Introduction}
\label{sectionI}

Trapped ions are considered highly promising for quantum computing due to their impressive characteristics, including long coherence time \cite{wang2017ten-minute,wang2021one-hour}, high operation fidelity \cite{Ballance2016QG-Ca,Gaebler2016QG-Be,srinivas2021Laser-free}, and full connectivity \cite{linke2017comparison}. 
To scale up the trapped ions and reduce the complexity of manipulation, two approaches are proposed for a stand-alone system. Firstly, the confinement potential can be engineered to be anharmonic to allow for the approximate uniform arrangement of a large number of ions \cite{Lin2009,Xie2017}. Alternatively, the Quantum Charge-Coupled Device (QCCD) architecture \cite{kielpinski2002architecture} can be implemented using surface electrode traps (SET) \cite{Chiaverini2005SET,Seidelin2006SET}. This approach involves controlling voltages to enable ions to shuttle between different functional zones segmented by multiple direct current (DC) electrodes, facilitating their interaction \cite{Jonathan2009QCCD,pino2021demonstration}. In both of these schemes, precise calibration of the spatial confinement potentials created by DC electrodes is essential. Particularly in the QCCD architecture, maintaining trap frequencies \cite{furst2014controlling} and minimizing phonon excitation \cite{Walther2012Transport,Bowler2012Transport,Ruzic2022} or creating trajectory with motional squeezing \cite{Sutherland2021Squeezing} necessitates precisely pre-calibration of the ion trap potential.

The SET with segmented shuttling-control electrodes offers an ideal platform for realizing the QCCD architecture. Various methods for potential estimation have been developed to optimize trap geometry and determine the best operating voltages. For instance, analytical methods have been established for calculating the potential of planar electrodes \cite{Oliveira2001BSL, Wesenberg2008Els}, with specific formulas derived for planar rectangular electrodes \cite{House2008Analytic}. While these methods offer convenience in trap design, their accuracy is limited due to the idealized assumption that vacancies in the electrode plane are filled by zero-potential electrodes, and ignored the presence of gaps between electrodes. Furthermore, these methods cannot handle the 3D structure of real traps, including finite metal thicknesses, multiple metal layers, oxide layers and loading holes. To overcome these limitations, numerical simulations, such as the finite element method (FEM) and the boundary element method (BEM) \cite{Singer2010RMP} are employed. 
However, even with numerical simulations, accurately replicating the true potential of an ion trap remains a significant challenge, primarily due to the presence of unexpected electrode defects, wire bonds, and environmental potentials induced by nearby entities in a real experimental system.

As expected, direct measurement of the actual potential is the most accurate method. Trapped ions can serve as excellent field probes for AC fields \cite{biercuk2010ultrasensitive,Gilmore2021,Feng2023}, DC fields \cite{berkeland1998minimization,harlander2010trapped,Feng2022}, and electrical noise \cite{Brownnutt2015RMP}. Measuring the trap frequency by shuttling an ion along the trap axis has been applied to probe the local field \cite{huber2010trapped}. Additionally, the spatial distribution of potential can be derived by measuring the equilibrium position of ions in a string by taking images \cite{brownnutt2012spatially}. These two methods focus on local and spatial electric fields, respectively, and the full trap potential information should be obtained by combining them. However, achieving this is currently challenging due to the intrinsic connection between these two aspects. For instance, in the spatial field measurement method \cite{brownnutt2012spatially}, large spacing between ions is preferred to enhance the measurement sensitivity, but this lead to larger interpolation errors \cite{brownnutt2012spatially}. Consequently, the trap frequencies calculated from the derived electric field is noisy, as it relates to the derivative of electric field strength and sensitive to local field fluctuations. Additionally, as the measured data are image-based, measurement accuracy is greatly affected by the quality of magnification calibration in the imaging system. Furthermore, during the calibration of the potential of a SET, the top-down asymmetry of the surface electrode structure typically leads to variations in the trapping height of ions in response to changes in electrode voltages. This can displace ion probes from their expected positions and introduce a excess systematic error.

In this paper, we introduce an optimization-based trap characterization method designed to reduce the systematic errors mentioned above, which also yields the benefit of obtaining a smooth and accurate spatial potential.
This method is based on the premise that the axial electrode potential can be expressed using a parametric empirical expression, and the stray field exhibits a simple form. The validity of the former is supported by BEM simulation results, and the latter is typically true for a limited trap region and a relatively stable trap environment during the experimental cycle. Compared to existing methods that rely on linear ion crystals \cite{brownnutt2012spatially}, our approach employs numerical optimization rather than interpolation and differential procedures, therefore reduces numerical errors introduced by interpolation and integration. Additionally, utilizing a multi-objective optimization method with constraints \cite{zhang2020versatile} allows us to exploit measured trap frequencies as auxiliary data, helping to minimize systematic errors resulting from magnification calibration error. Most importantly, in contrast to previous method which requires changing the voltage on each pair of electrodes in a regular interval and alternating manner, our method allows more flexible voltage settings to move ion string probes, which can maintain constant trapping heights in a SET and reducing the systematic error caused by ion height variations. The accuracy of our new method has been validated through a comprehensive comparison with the existing one, conducted under nearly identical experimental conditions.

The structure of the paper is as follows: Section \ref{sectionII} provides a review of existing trap characterization methods and presents the principles underlying our optimization-based approach. In Section \ref{sectionIII}, we describe the experimental setup for data collection. The primary results of this study are presented in Section \ref{sectionIV}, where we obtain the field strengths of electrodes and ambient sources using both methods and compare them with experimental data. Finally, we conclude in Section \ref{sectionV}.

\section{Theory}
\label{sectionII}
\subsection{Brief Review of the Existing Trap Characterizing Methods }

The linear SET utilizes radiofrequency (RF) electrodes for radial confinement and DC electrodes for axial confinement and shuttling control. In our SET setup, all electrodes are nearly rectangular and situated in the same plane. 
The unit-voltage potential $\phi_{k}$ is generated when a voltage of 1 V is applied to the $k$-th electrode while all other electrodes remain at 0 V. The total axial potential in the SET, excluding the RF pseudopotential component, can be expressed as:
\begin{equation}
	\Phi_t = \sum_{k =1}^{N}V_{k}\phi_{k},
\end{equation} 
where, $N$ represents the number of DC electrodes, and $V_{k}$ is the voltage applied to the $k$-th electrode. Consequently, accurately determining the form of $\phi_{k}$ is the primary goal of trap potential characterization. Additionally, there exists a static ambient potential stemming from such as unexpected dielectric near the trap, which we label as $\Phi_s$ and also aim to determine in the following analysis.

We begin by reviewing the theoretical methods for describing trap potentials $\Phi_t$. The electrostatic potential of a rectangular electrode can be calculated analytically by the method developed by M. G. House \cite{House2008Analytic}. This model assumes that the electrodes extend infinitely within the plane and have infinite small gaps, and the static potential of a unit-voltage rectangular electrode is then described by the following expression:
\begin{eqnarray} 
	\label{AnaRec}
	\phi_{k}(x,y,z)=&\frac{1}{2\pi}\bigg\{ \arctan\left[\frac{(x_{k2}-x)(z_{k2}-z)}{y\sqrt{y^{2}+(x_{k2}-x)^{2}+(z_{k2}-z)^{2}}}\right] \notag \\ 
	&-\arctan\left[\frac{(x_{k1}-x)(z_{k2}-z)}{y\sqrt{y^{2}+(x_{k1}-x)^{2}+(z_{k2}-z)^{2}}}\right] \notag \\
	&-\arctan\left[\frac{(x_{k2}-x)(z_{k1}-z)}{y\sqrt{y^{2}+(x_{k2}-x)^{2}+(z_{k1}-z)^{2}}}\right] \notag \\
	&+\arctan\left[\frac{(x_{k1}-x)(z_{k1}-z)}{y\sqrt{y^{2}+(x_{k1}-x)^{2}+(z_{k1}-z)^{2}}}\right] \bigg\},
\end{eqnarray}
here, $(x_{k1}, 0, z_{k1})$ and $(x_{k2}, 0, z_{k2})$ represent the coordinates of opposite corners of the $k$-th electrode.

However, it's important to note that this expression does not precisely match the actual potential. A more accurate unit-voltage potential $\phi_{k}$ can be obtained by numerical simulations employing the standard BEM or FEM. 

While these theoretical approaches are valuable in many aspects, they exhibit limitations when it comes to deriving $\Phi_s$, and do not meet the required accuracy for precise shuttling control. Therefore, we are pursuing a measurement method to determine the unit-voltage potential $\phi_{k}$ and stray field $\Phi_s$.

We will now provide a concise overview of the measurement method as demonstrated by M. Brownnutt \textit{et al.} \cite{brownnutt2012spatially}. For simplicity, we consider single-charged ions confined to one dimension (1D) along the x-axis, subjected to a confining potential $\Phi_t+\Phi_s$. Each ion, denoted as $i$ and positioned at $x_{i}$ within the stationary linear chain, experiences a Coulomb repulsion force originating from all other ions, as described by the following equation:

\begin{equation}
	\label{Fion}
	F_{ion}(x_i) = \frac{e^{2}}{4\pi \epsilon_{0}} \sum_{j \neq i}^{} \frac{|x_{i}-x_{j}|}{(x_{i}-x_{j})^{3}}.
\end{equation}

In this equation, $e$ represents the charge of an ion, $\epsilon_{0}$ denotes the vacuum permittivity, while $x_{i}$ and $x_{j}$ signify the positions of ions $i$ and $j$, respectively. This force is equal in magnitude and acts in opposition to the external force generated by the confining potential, which is associated with the electric field intensity $E_{ext}(x_i)=-F_{ion}(x_i)/e$.

\begin{figure*}[thbp]
	\centering
	\includegraphics[width=0.65\linewidth,scale=1.000]{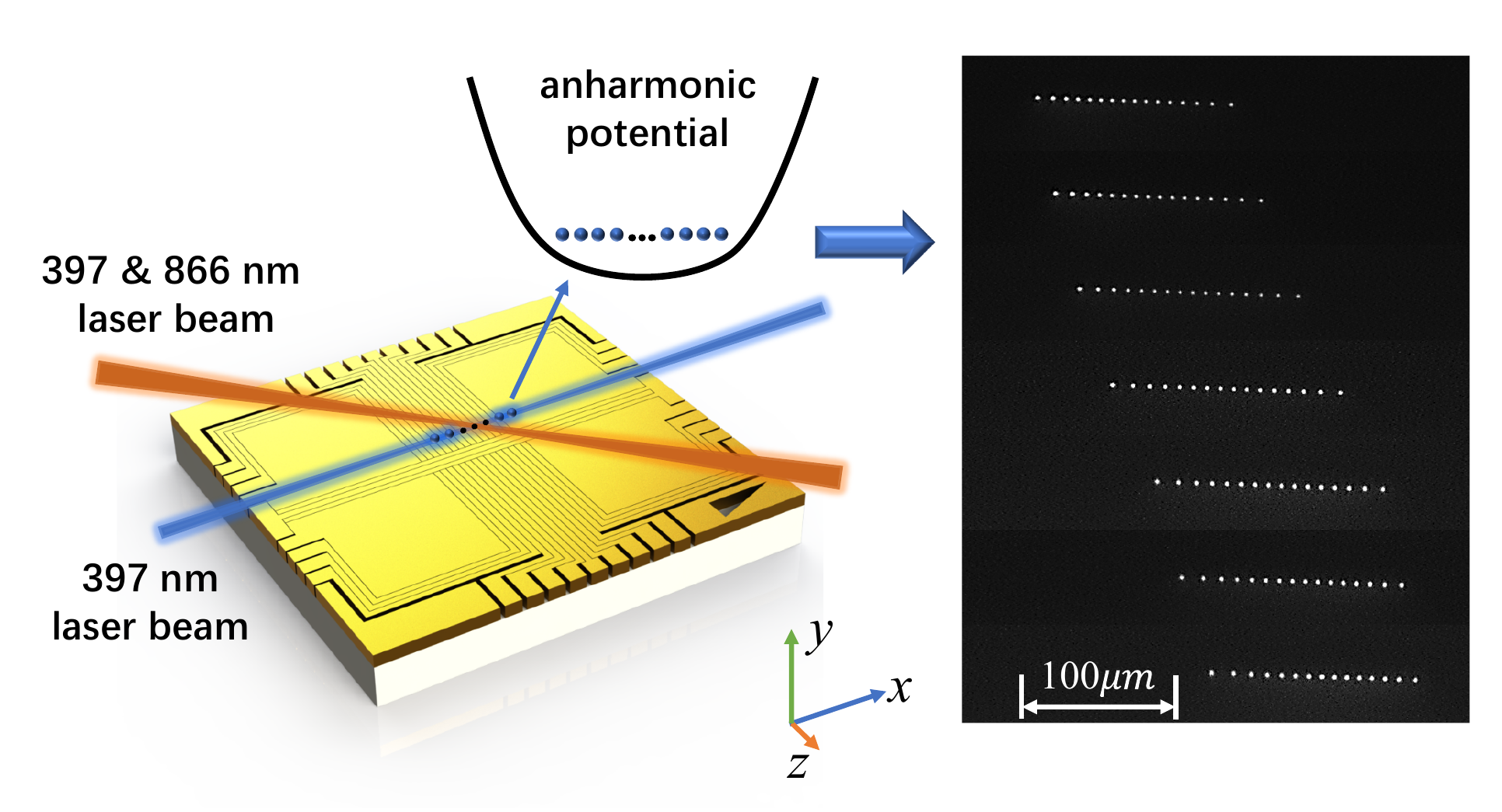}%
	\caption{\label{Fig 1}{Ion string used as the potential probe. An linear ion chain is trapped above the SET and Doppler-cooled by the 397 nm and 866 nm laser light. Every time the voltage on one specific pair of the DC electrodes changed, the ion string will move to a new equilibrium position. The moving ion strings allow us to probe the spatial field distribution.}}
\end{figure*}

Using the ion positions as interpolation points, the function $E_{ext}(x_i)$ can be numerically integrated to determine the instantaneous confining potential in 1D, albeit with an unimportant unknown integration constant. It's worth noting that the force $E_{ext}(x_i)$ comprises two components: $E_{ext}(x_i) = E_{t}(x_i) + E_{s}(x_i)$. Here, $E_{t}(x_i)$ is associated with all the DC electrodes and is voltage-dependent, while $E_{s}(x_i)$ arises from all other unspecified sources and remains voltage-independent, often referred to as the stray field. The corresponding potentials are $\Phi_t$ and $\Phi_s$, respectively.

To measure the unit-voltage potential $\phi_{k}$, the voltage on the $k$-th pair of electrodes is systematically varied  by a small increment $\delta$. The total potential $\Phi=\Phi_t+\Phi_s$ depends not only on the voltage applied to the electrode of interest, $V_{k}$, but also on the voltages of the other electrodes collectively referred to as $V_{B}$. The unit-voltage potential provided by the electrode of interest can be calculated by the expression:
\begin{align}
	\phi_k(x) & =\Phi_t(x,V_{k} = 1, V_{B} = 0 )\\ 
	          & = \left[\Phi(x, V_{k}+\delta, V_{B}) - \Phi(x,V_{k}, V_{B})\right] \times 1 V /\delta \notag .
\end{align} 

Since the change in voltage $\delta$ result in a shift in the ion positions  $x_i$, it is possible to compute the potential $\Phi_t(x, 1, 0)$ for all values of $x$ within the overlapping range of the two datasets, $\Phi(x, V_A + \delta, V_B)$ and $\Phi(x, V_A, V_B)$, with the aid of data interpolation. The measurable regions can be expanded as the ion string moves with adjusted $V_k$.

Error analysis has indicated that the accuracy of this method is limited by data interpolation. The measurement uncertainty of the field is smaller when ion densities are lower, however, the numerical interpolation become less accurate in the limit of low ion density. Yet, even when averaged, the smoothness of the curve $\Phi_t(x, 1, 0)$ is not guaranteed, which may lead to fluctuations when calculating the trap frequency.

Furthermore, it's crucial to acknowledge that while the demonstration is conducted in a 3D trap, as opposed to a SET, the trapping height remains unaffected even when altering only one pair of DC voltages at a time. However, this scenario differs in a SET, where changes in DC voltages usually lead to variations in trapping height. To maintain the trapping height as constant as possible in a SET, it becomes imperative to ensure that the vertical component of the electrode field strength remains consistent. This can be facilitated by calculating the trapping voltage settings using the 3D potential derived from numerical simulations. However, deriving the unit-voltage potential using this method in a SET is not as straightforward.

\subsection{Basic Theory of The Optimization-Based Method}

We present a data processing method rooted in numerical optimization with the objective of minimizing the disparity between the data predicted by the model and the experimental measurements. To circumvent the need for integration, we opt to directly determine the unit-voltage electric field intensity of the $k$-th electrode, denoted as $E_{k}(x, 1, 0)$, instead of $\phi_{k}(x, 1, 0)$. This approach necessitates the establishment of a parametric expression for the electric field intensity. In the case of rectangular electrodes, we may employ the partial derivatives of Eq. (\ref{AnaRec}). Here, the 1D distribution along the trap axis is obtained by setting $y$ as the trap height and $z=0$. However, this equation is excessively complex for optimization purposes.

Our investigation has revealed that the 1D unit-voltage potential curve along the x-axis, derived either from Eq. (\ref{AnaRec}) or the BEM, can be effectively approximated by an unnormalized Lorentz curve, with an error margin of only a few thousandths under specific conditions discussed in the Appendix. As depicted in Fig. \ref{LorentzModel}(a), the unit-voltage 1D potential of the 8-th electrode, calculated using the BEM method, closely matches a Lorentz function for a rectangular electrode with 147 $\mu$m wide and 0.94 mm long. Additionally, the axial component of the electric field strength closely matches the first derivative of the Lorentz function, as shown in Fig. \ref{LorentzModel}(b).

\begin{figure}[thbp]
	\centering
	\includegraphics[width=0.95\linewidth,scale=1.000]{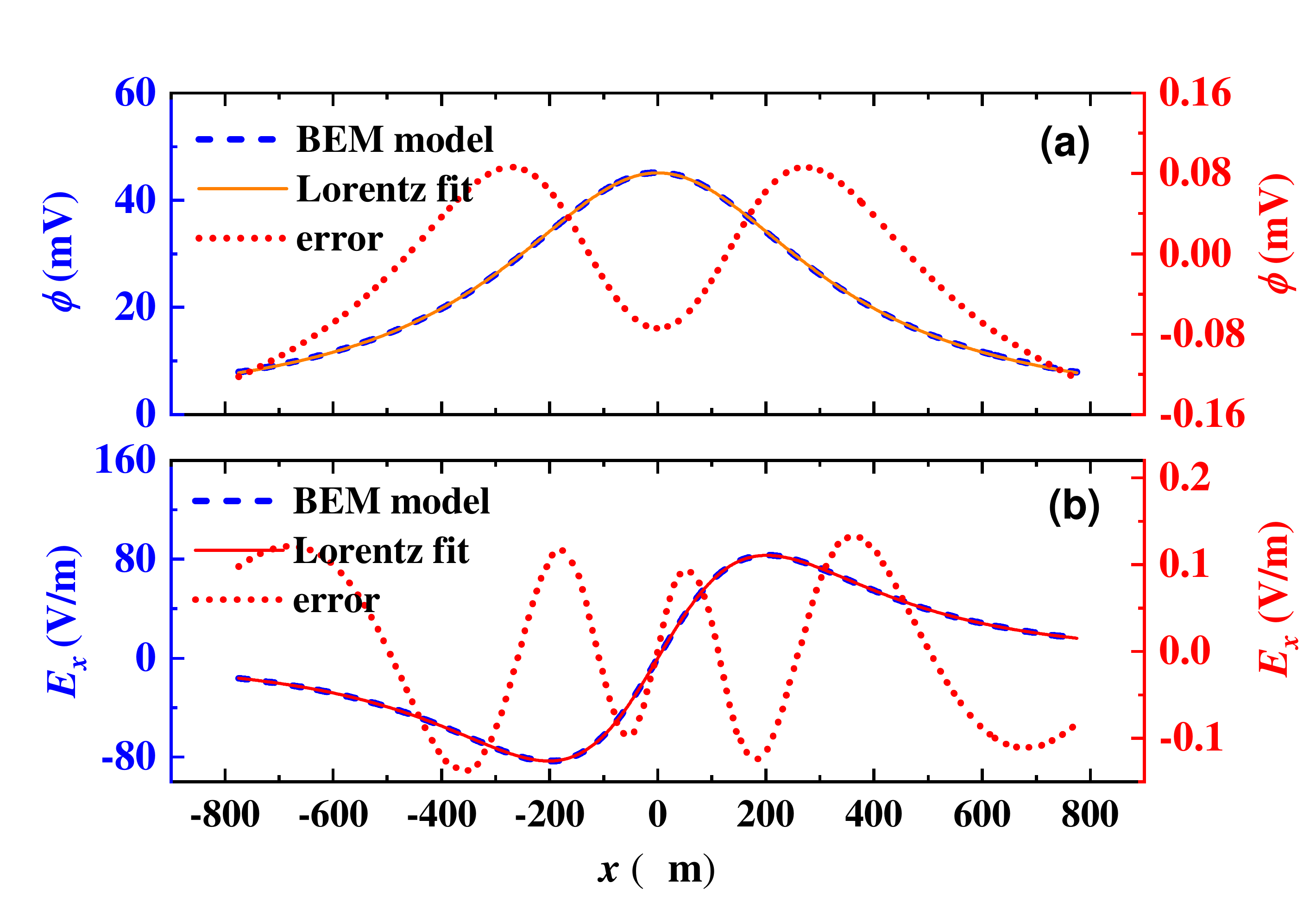}%
	\caption{\label{LorentzModel}{Lorentz fit of the 1D unit-voltage potential curve along trap axis. (a) The potential curve and (b) the axial component of the electric field strength. The blue dashed lines are calculated by BEM method, and the red solid lines are derived by the fitted Lorentz function. The fitting errors $\Delta \phi$ and $\Delta E_x$ are shown in red dotted line.}}
\end{figure}

Hence, we initiate the process by proposing an ansatz for the parametric expression of the electric potential associated with the $k$-th rectangular electrode:
\begin{equation}
	\label{Lorentz}
	\phi_{k}(x)= \frac{A_k \gamma_k}{(x-x_k)^2+\gamma_k^2}.
\end{equation} 
The free parameters, denoted as $A_k$ and $\gamma_k$, are to be determined, and $x_k$ represents the central position of the $k$-th electrode. Subsequently, the $x$-component of the electric field intensity can be computed as $E_{k}(x) = -\partial\phi_{k}(x)/\partial x$. These parametric functions are employed to express the $x$-component of the total electric field intensity as 
$E(x)=\sum_{k=1}^{N}V_{k}E_{k}(x)+E_s(x)$.

We assume that the stray electric field remains constant throughout the measurement process and the axial distribution of the stray electric field conforms to the following form:
\begin{equation}\label{Es}
	E_{s}(x)= a x^2 +b x +c,
\end{equation} 
where, $a$, $b$ and $c$ are yet to be determined. Please note that higher-order terms should be added if a long shuttling range or a complex environment are involved.

The optimization method's primary objective is to minimize the sum of squared errors between the predicted and measured trapping forces, (i.e. $eE_{ext}(x)$ and $-F_{ion}(x_i)$), for all ions in a string across different voltage settings. 

In addition to extracting information from the equilibrium positions of the ion string, we can utilize secular motion frequency data acquired at various voltage settings to determine model parameters. Unlike the equilibrium position of the trapped ions, the secular motion frequency is intricately connected to the second derivative of the potential at the location of the potential minimum, denoted as $D(x_i) = \sum_{k=1}^{N}V_{k}D_{k}(x_i) + D_s(x_i)$, which can be expressed as follows:
\begin{equation}\label{omegax}
	\omega_x= \sqrt{\frac{e D(x_i)}{M_{ion}} }.
\end{equation} 
here, $x_i$ represents the equilibrium position, with the notation $D_k(x_i)=\frac{\partial^2\phi_k(x)}{\partial x^2}\vert_{x=x_i}$,
$D_s(x_i)= \frac{\partial^2\Phi_s(x)}{\partial x^2}\vert_{x=x_i}$.
This secular frequency is unrelated to ion imaging, and systematic errors associated with magnification calibration can be partially addressed by taking it into account.

By leveraging different datasets with distinct characteristics, we can enhance the precision of local field determination. To maximize the utility of all these datasets, the characterization process can be viewed as a multi-objective optimization problem with two objective functions:

Furthermore, the equilibrium position ($x_i$) of a singular ion, trapped under specific voltage conditions, acts as a constraint in our solution. This position is ascertained within the trap model by locating the root of $E(x_i) = 0$. While the dataset from a single ion provides high accuracy, primarily only influenced by the positional uncertainty of the ion itself, it possesses limited spatial coverage compared to the ion string dataset.

By leveraging diverse datasets, each exhibiting unique characteristics, we can improve the precision of local field determination. To optimize the use of these datasets, the characterization process can be conceptualized as a multi-objective optimization problem with two objective functions:
\begin{eqnarray}
t_{1}= \sum_{i,j}^{}|\widetilde{E}_{ion}(U_{j},x_{j,i})-E_{s}(x_{j,i})-
\sum_{k=1}^{N}V_{j,k}E_{k}(x_{j,i})|^{2} \notag \\
t_{2}= \sum_{j}^{}\bigg|\widetilde{\omega}_{x}(U_{j},x_{j})-\sqrt{\frac{eD_{s}(x_{j})}{M}+\sum_{k=1}^{N}\frac{eV_{j,k}D_{k}(x_{j})}{M}}\bigg|^{2}  \notag
\end{eqnarray}
subject to \qquad $|E_{s}(x_j)+\sum_{k=1}^{N}V_{j,k}E_{k}(x_j)|\le  \Delta \widetilde{E} $.

In this formulation, $x_{j,i}$ (or simply $x_j$ for a single ion) represents the position of the $i$-th ion in a linear chain under the $j$-th voltage setting $U_{j}$, where the $k$-th electrode has a voltage of $V_{j,k}$. $\widetilde E_{ion}(U_j,x_{j,i})$ and $\widetilde{\omega}_{x}(U_j,x_j)$ denote the measured electric field intensity with ion strings and measured secular frequency, respectively, under specific voltage settings $U_j$. Here, $N$ signifies the total number of electrodes. The undetermined parameters, $A_k$ and $\gamma_k$, are embedded within the expressions of $E_s$, $E_k$, $D_s$, and $D_k$. The constraint is designed to align the predicted position of a single ion with the measured $x_i$, while considering a field intensity uncertainty $\Delta \widetilde{E}=M\widetilde{\omega}_x^2\Delta x /e$ arising from the random error in the ion's position ($\Delta x$).

The total number of undetermined parameters in this scenario amounts to $2N+3$, growing linearly with the number of electrodes involved. For instances involving fewer than 10 electrode pairs, as demonstrated in this work, the problem can be effectively tackled using global optimization algorithms like Differential Evolution. In scenarios with a higher number of electrodes, we suggest dividing them into multiple groups, each capable of trapping ions and independently characterized experimentally. This approach will enhance the efficiency of the optimization process, shorten the experimental duration, and aligns well with the assumption of a constant stray field.

Our method eliminates the need for data interpolation since it focus on sampling points that coincide with ion positions. Moreover, our approach combines the advantages of analytical functions and experimental measurements, ensuring a seamless and precise potential. The optimization algorithm is adaptable to various types of experimental data, thereby enhancing the model's accuracy. Furthermore, this method is well-suited for SET as it enables the control of the ion string's movement by adjusting DC voltages applied to all the electrodes collectively, rather than one pair at a time. This approach allows for the maintenance of the trapping height throughout the entire measurement through pre-voltage design. The advanced voltage update strategy, however, is depend on the 3D potential of the trap model and beyond of the scope of this work.

\section{Experimental Demonstration}
\label{sectionIII}

\begin{figure}[thbp]
	\centering
	\includegraphics[width=0.98\linewidth,scale=1.000]{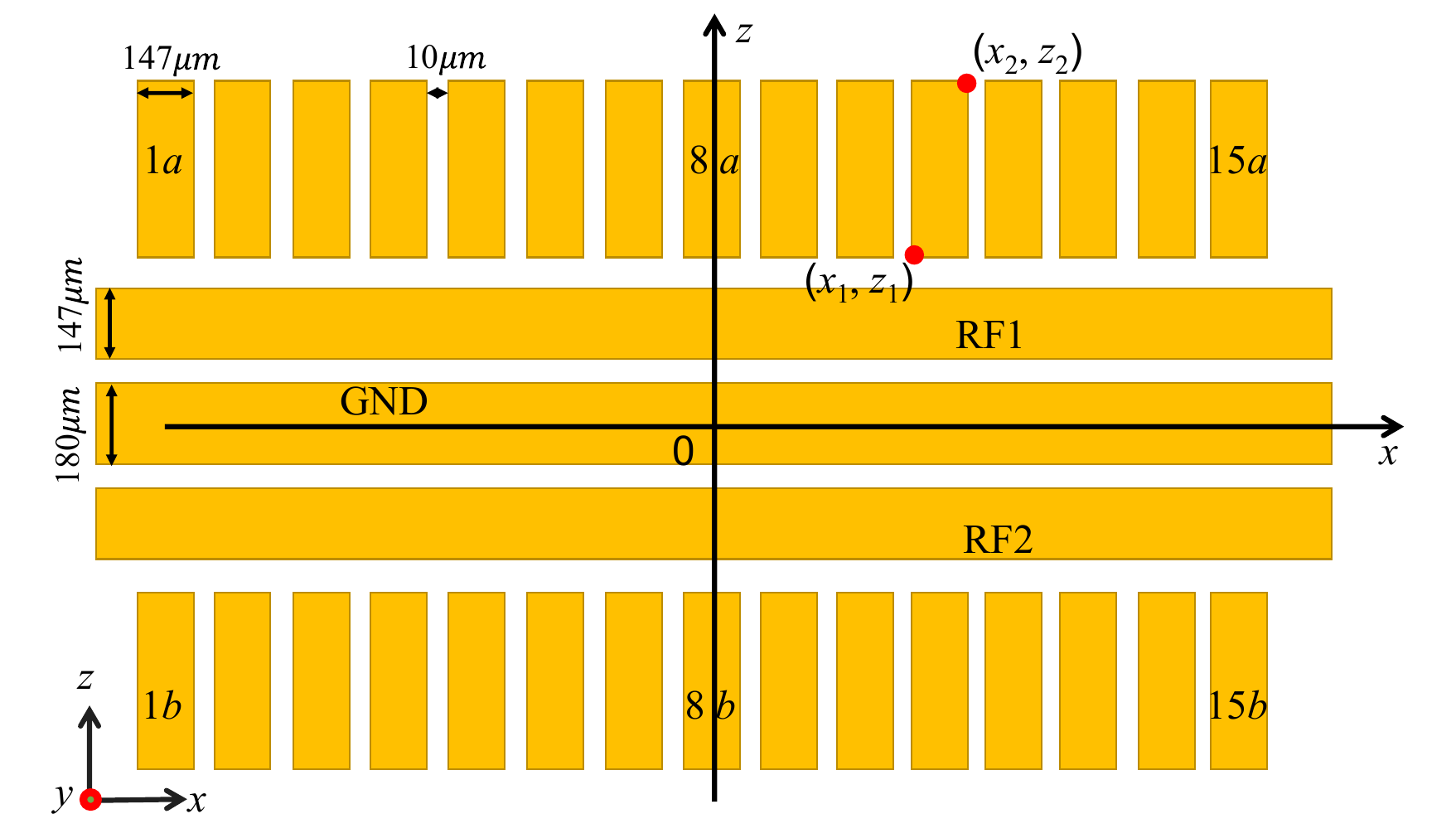}%
	\caption{\label{Fig 2}{Schematic diagram of the "five wire" linear ion trap. The DC electrodes of 147 $\mu$m located at the $y=0$ plane are labeled from $1a(b)$ to $15a(b)$}. The RF electrodes (RF1(2) ) have the width of 147 $\mu$m and GND electrode is 180 $\mu$m wide.}
\end{figure}

We implement the scheme in our "five wire" linear SET. The apparatus is described in reference \cite{ou2016optimization}. The trap consists of fifteen pairs of DC electrodes, as shown in Fig. \ref{Fig 2}. 
The DC electrodes labeled from $1a(b)$ to $15a(b)$ are used for axial confinement. The other electrodes (RF1(2) and GND) provide the transverse confinement. The RF loaded into the trap has a frequency of $\Omega_{rf}/2\pi=22.7$ MHz, resulting in a transverse trap frequency of about 2.6 MHz. 
The tight confinement achieved allows us to move the ion crystal along the trap axis without altering the trapping height too much when adjusting a single pair of DC electrodes. To mitigate height variations, an improved voltage updating strategy is essential. However, for the purpose of this work, we opted to vary the voltage of a single pair of electrodes to compare results with the existing method. The associated systematic error is then assessed through simulated 3D potential.

Our axial confining potential is powered by 9 channels of a 16-bit resolution DAC device, offering an output range of $(-10V\sim 10V)$. Specifically, only the central nine (i.e. $4a(b)$ to $12a(b)$) out of the fifteen pairs of DC electrodes are actively used, with the remaining pairs grounded.  

The $^{40}\text{Ca}^{+}$ ions are loaded through a three-step photo-ionization process from Ca atoms, which are evaporated by heating an atom oven. This loading method utilizes laser light with wavelengths of 423 nm and 732 nm, as described in \cite{zhang2017realizing}. A linear chain of the $^{40}\text{Ca}^{+}$ ions is confined in an anharmonic potential along the trap axis. The minimum spacing between adjacent ions is above 10 $\mu$m to ensure the field sensing sensitivity. The linear chain is Doppler cooled with 397 nm and 866 nm laser light. We have two 397 nm laser beams, one is along the $(1,0,0)$ direction which provides cooling in the $x$ direction. The other is slightly tilted away from the $(1,0,1)$ direction, predominantly providing cooling in the $z$ and $x$ direction, while having only a little component in the $y$ direction. The uncooled motion in $y$ direction is basically irrelevant in this experiment, since the sensitive surface of the camera is oriented perpendicular to this direction. 

Cooling ions to the Doppler-limit is not necessary, but suppressing the micromotion as much as possible is important, because this will reduce the uncertainty of ion position. When the voltages are identically applied to the pair of electrodes, micromotion is negligible in the $z$ direction in our trap. Coarse micromotion compensation in the $y$ direction is achieved by adjusting the height of the ions above the trap until the images of the individual ions are best localized on the camera. The number of ions in the chain will decrease gradually due to the collision with residual gas in the vacuum chamber. Loading process will be launched to keep the ion number within 6 to 19 in an experiment.

In the experimental scheme, each pair of the electrodes labeled "$i$a" and "$i$b" ($4 \le i \le 12$) are loaded with the same voltages respectively, and thus the unit-voltage potentials are determined by pairs. In the ion string data set acquisition stage, voltage on each $i^{th}$ pair are repeatedly updated with a voltage increment of $\delta_{i}\sim 0.02V$ while keeping all the other voltages unchanged, pushing the ion string move across the region of interest ($\sim 280\mu$m, limited by the beam width of diagonal 397-nm laser, and within the fitting range of the Lorentz function). 

Note that keeping the $\delta_{i}$ constant for a specific $i^{th}$ electrode and the other voltages constant is necessary for the interpolation method, but not for ours. Actually, to cover wider operating voltage range and mitigate systematic error caused by trapping height variation, change voltages on different electrodes simultaneously is preferred. As a contrast, recording all the voltages on each electrode is necessary for the latter but not for the former. We keep the $\delta_{i}$ constant in the experiment, ensuring that the outcomes of both methods can be effectively compared using a uniform dataset. In this way, the efficiency of the new method can be verified. Concurrently, we evaluate the impact of height variation on systematic errors by employing a simulated 3D potential model.

Every time the voltages of the DC electrodes are updated, an image of the linear ion string is taken by an electron-multiplying charge-coupled device (EMCCD iXon Ultra 888). The customized lens provides approximately 19.2x magnification, resulting in a final imaging range of $693~\mu m$ by $693~\mu m$, with a resolution of $0.676~\mu m$ per pixel size. The magnification of the system is calibrated by capturing the image of a trap electrode with a known width. This calibration is further verified through the imaging of two trapped ions, allowing for the precise calculation of their separation based on the measured trap frequency.

The position of an ion in the string under voltages $U_j$ is determined using a 2D Gaussian fit. For each ion $i$, we first derive the center of mass position. Then, the image is partitioned into sections from the midpoint between two adjacent ions, with each section containing only one ion, and the 2D Gaussian fit is applicable. The position error, estimated to be less than 0.4 $\mu$m, is determined by the fitting quality. These positions $x_{j,i}$ are then utilized to calculate the electric field intensity $\widetilde{E}_{ext}(U_{j},x_{j,i})$ using Eq. (\ref{Fion}).

The procedure for acquiring the equilibrium position of a single ion $x_j$ under certain voltages $U_j$ is similar. Upon adjusting the voltage settings, the image of a single trapped ion is captured, and a 2D Gaussian fit is applied to determine the ion's position. Throughout the measurement, both the exact position of the ion trap and the imaging system remain unchanged.

The secular frequencies $\widetilde{\omega}_{x}(U_j,x_{j}) $ are then measured by resonant excitation, with the equilibrium positions and voltage settings recorded at the same time. The excitation signal provided by a sine wave generator is connected to the outermost DC electrode. To ensure utmost accuracy a single trapped ion and a very weak resonant excitation signal are employed. Fluorescence levels change as the excitation frequency sweeps across the resonance point. The measurement uncertainty is maintained at less than $\pm 0.5$ kHz. When acquiring different types of data, the voltage settings don't need to be the same.

In our experiment, the number of undetermined parameters is as large as 21. Consequently, the datasets must be sufficiently extensive to minimize parameter uncertainty. We capture over 30 images for each pair of electrodes under varying voltages. Each image comprises 6 to 19 ions, along with 20 secular frequencies and positions of two single-trapped ions (more would be preferable). The total number of data points exceeds 3500, all of which are utilized in our optimization method. However, the interpolation method can only make use of a subset of these data points. Position data near the ends of the ion string is not useful for the interpolation method, as the number of overlapped samples is insufficient for averaging to effectively reduce random errors.

For comparison, both the interpolation method and our optimization approach are employed to derive the unit-voltage field intensity for each pair of DC electrodes, as well as for the stray field.

\section{Result}
\label{sectionIV}
~\\
\textbf{\emph{Electric field intensity of the electrodes and stray field}}

\begin{figure}[thbp]
	\centering
	\includegraphics[width=1\linewidth,scale=1.000]{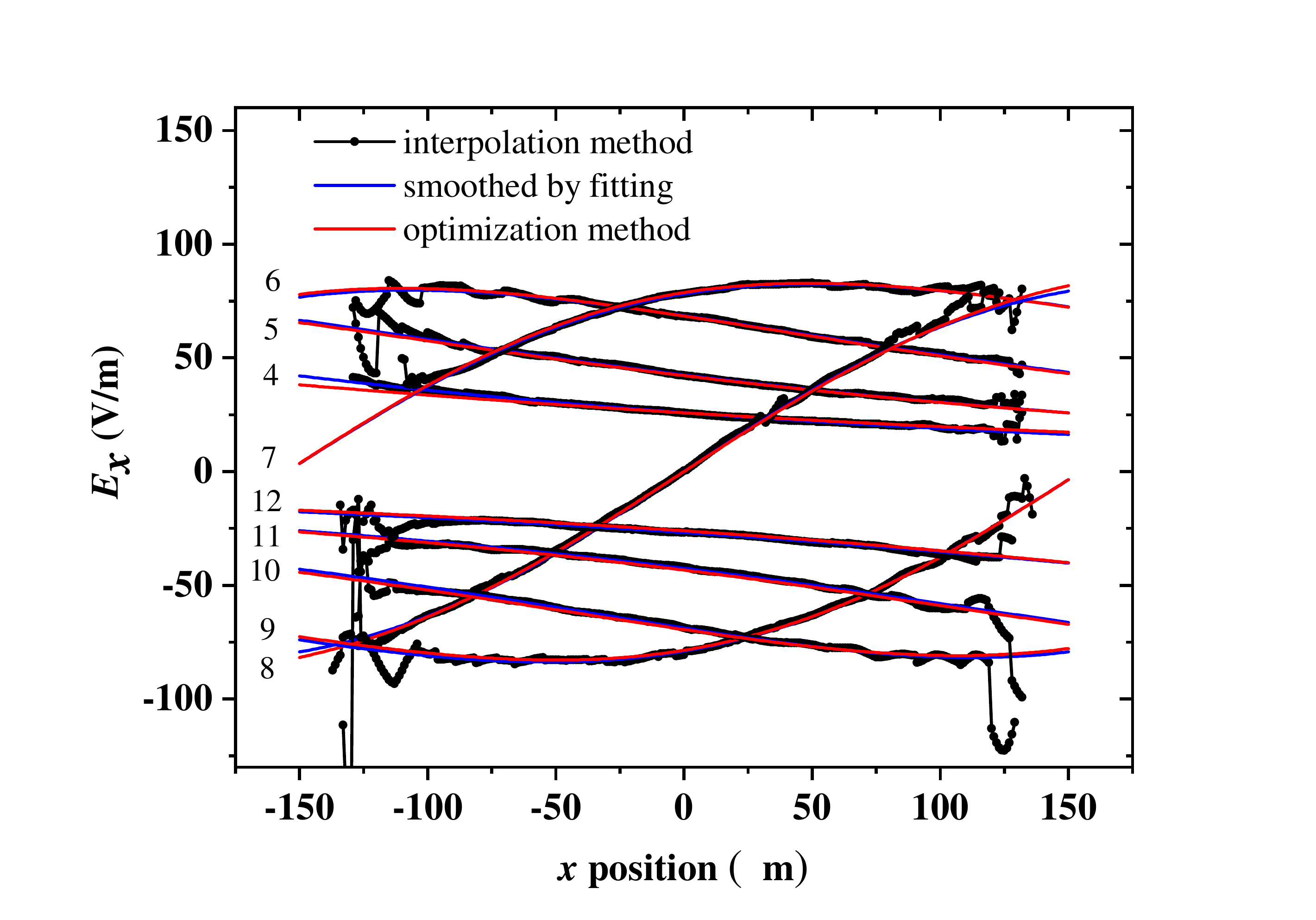}%
	\caption{\label{FieldStrength}{ Unit-voltage electric field intensity curves of electrode $4-12$ derived by different methods. Each curve corresponds to the $i^{th}$ electrode, labeled by number $i$ on the left. The black solid lines with dots represent results from the interpolation method. The discrete data within the range $(-110,110)\mu$m are fitted using the Lorentz function Eq. (\ref{Lorentz}), resulting in the blue solid lines. The red solid lines are derived using the optimization method, utilizing all experimental data.
	}}
\end{figure}

We first use the interpolation method proposed by M. Brownnutt \textit{et al.}\cite{brownnutt2012spatially} to calculate unit-voltage electric field intensity of the DC electrodes in pairs, utilizing only the ion string data set. The results, depicted as black dotted lines in Fig. \ref{FieldStrength}, exhibit noticeable random fluctuations, particularly at the two ends of the region where the samples for averaging are scarce. To characterize the trap better, we smooth these curves by fitting them with a Lorentz function based on Eq. (\ref{Lorentz}), but restrict the fit to data within $-110 \sim 110 \mu $m to avoid significant errors, as illustrated by the blue lines in Fig. \ref{FieldStrength}. 

In the same figure, the red lines represent the results obtained using our newly proposed optimization method, which utilizes all collected data without discarding the ends. The optimization targets, $t_1$ and $t_2$, are combined and balanced with a weighting factor. Two positions of single-trapped ions under different voltages are used to constrain the solution.

\begin{figure}[thbp]
	\centering
	\includegraphics[width=0.95\linewidth,scale=1.000]{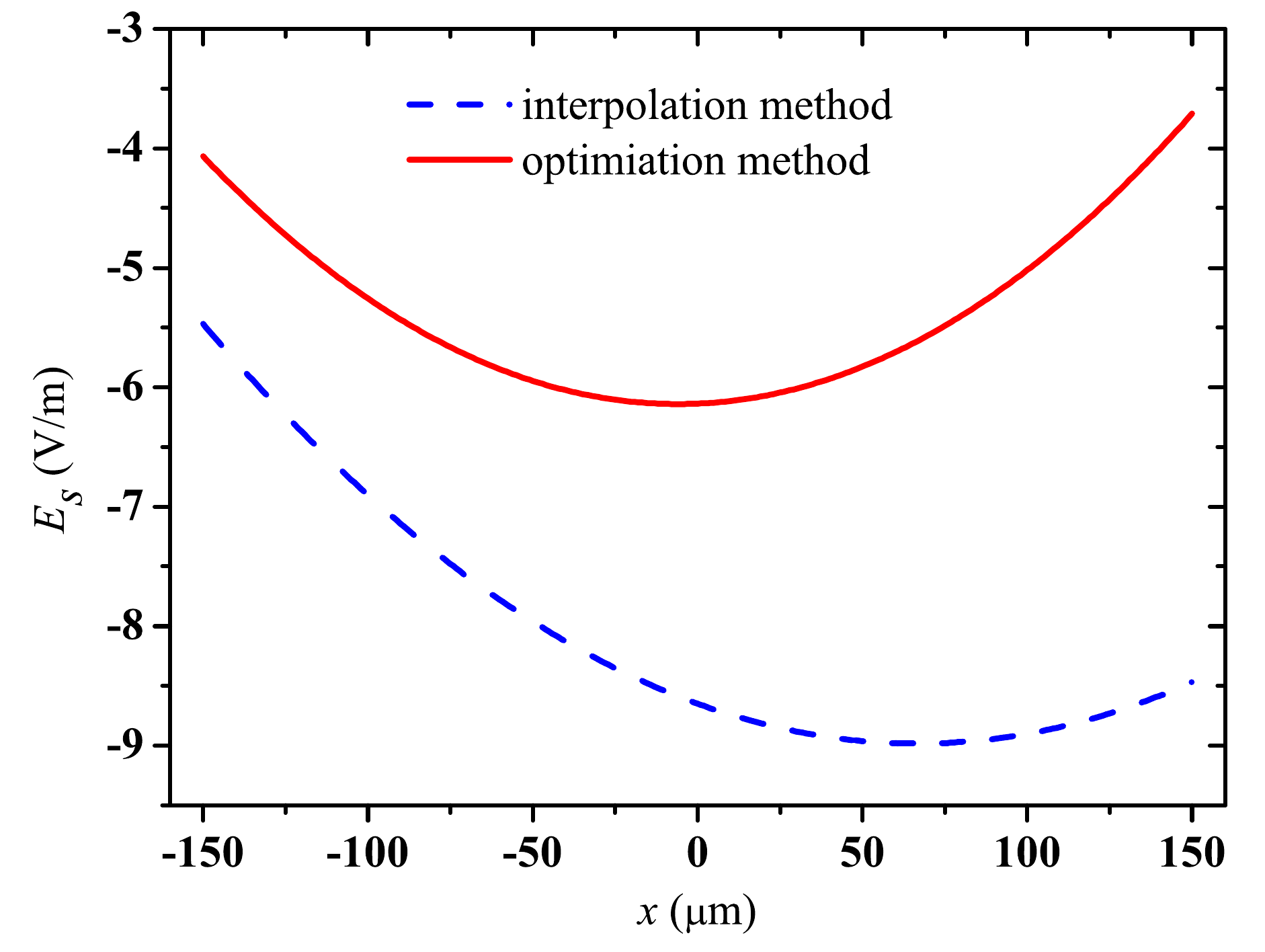}%
	\caption{\label{StrayField}{ The stray electric field intensity $E_s$. The blue dashed line (red solid line) is derived using the trap model according to interpolation (optimization) method. }}
\end{figure}

The stray electric field can be derived using both methods. In the optimization method, it is solved directly. However, in the interpolation method, the stray electric field has been subtracted as a background. Therefore, we calculate the residual error between the measured electric field and the predicted one after deriving all the unit-voltage electric field intensities. The stray electric field intensity along the axis is separately derived by the two methods, as shown in Fig. \ref{StrayField}. As evident from the results presented in the following section, the frequencies and ion position errors calculated by our method are smaller. Therefore, we can infer that the stray fields obtained by our method better match reality. We hypothesize that this advantage is primarily attributed to the reduction of systematic errors and interpolation errors. The curves indicate that the main source of the stray field is not far from the trap center, possibly originating from the light charging effect due to the laser beams.

The unit-voltage field intensity of each electrode in the blue (red) curves in Fig. \ref{FieldStrength}, multiplied by the applied voltages, along with the stray field represented by the blue (red) curve in Fig. \ref{StrayField}, constitute the trapping field $E(x)$. The calibrated model $E(x)$ can then be used to simulate trapping dynamics.

\begin{figure}[thbp]
	\centering
	\includegraphics[width=0.95\linewidth,scale=1.000]{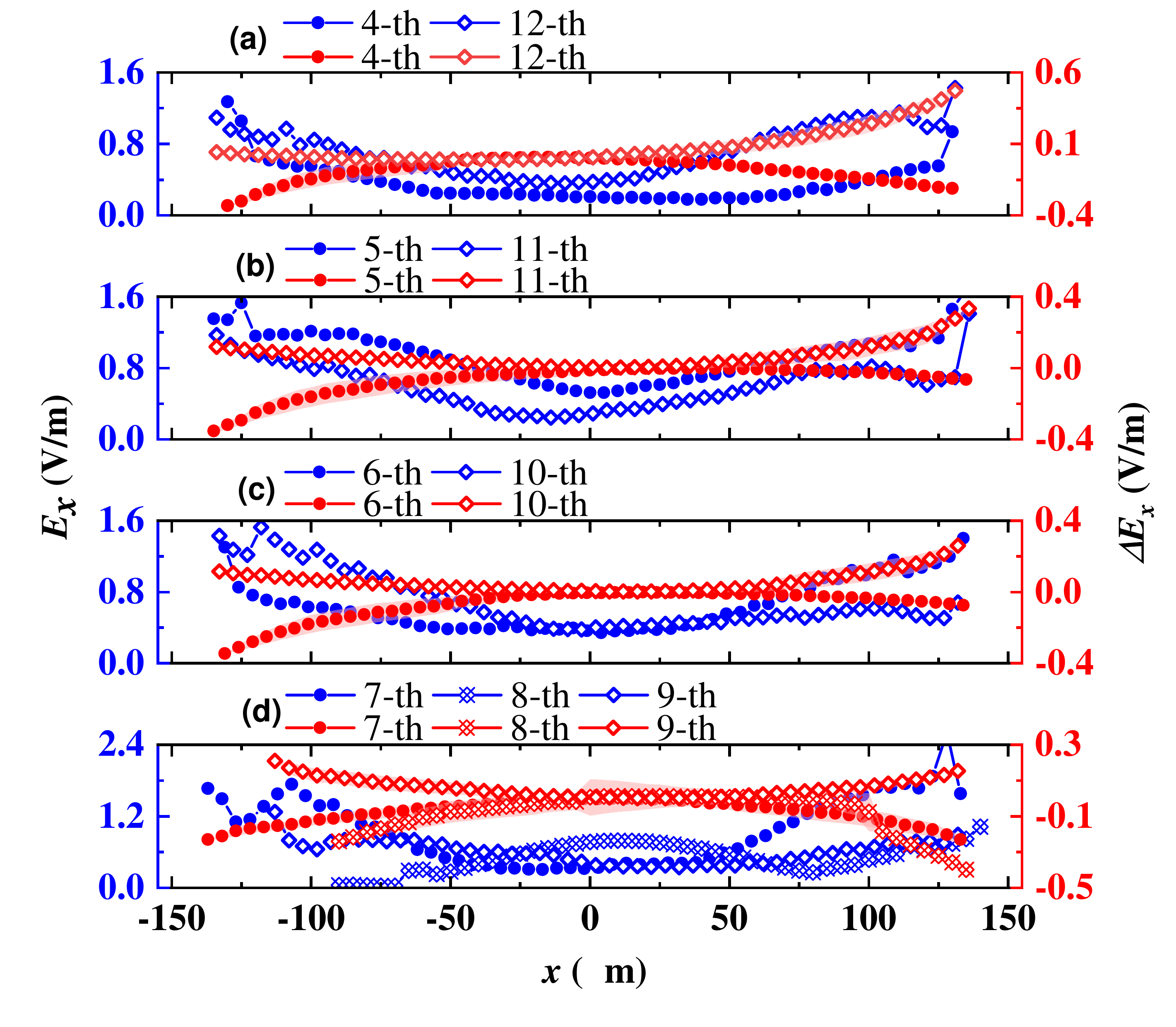}%
	\caption{\label{RandomErrAndSysErr}{ Measurement error of the ion string field probe. The random error $\delta E_x$ induced by imaging resolution is shown in blue, and the systematic error $\Delta E_x$ induced by ion height variation is shown in red. The error associated with different voltages on certain electrodes are grouped in to sub-pictures: (a) 4-th (in circle) and 12-th (in diamond) electrodes, (b) 5-th (in circle) and 11-th (in diamond) electrodes, (c) 6-th (in circle) and 10-th (in diamond) electrodes, (d) 7-th (in circle), 8-th (in X-marked diamond) and 9-th (in diamond) electrodes. }}
\end{figure}

The random errors of the measured potential $\delta E_x$, are primarily constrained by the resolution of the imaging system and is independent of the data processing method. These random errors vary depending on relevant positions and electrodes, as is shown in the blue marks of Fig. \ref{RandomErrAndSysErr}. These calculations consider all the random errors of ion positions in the string and contributions of all the involved ion string by $\frac{1}{N_{sample}}\sum_{j=1}^{N_{sample}}\sum_{i=1}^{N_{ion}}(E_{ion}(x+\delta x)-E_{ion}(x))^2$. It is evident that, except for those associated with the 8th pair of electrodes, the random error is significantly greater for positions at the ends compared to those in the middle, due to the much lower number of overlapped images in these regions. The reason why the 8th pair of electrodes is an exception is that it has a very weak $E_x$ component. As a result, ions are less sensitive to changes in the electric field strength in this region, leading to a relatively small overall random error. 

\begin{figure}[thbp]
	\centering
	\includegraphics[width=0.95\linewidth,scale=1.000]{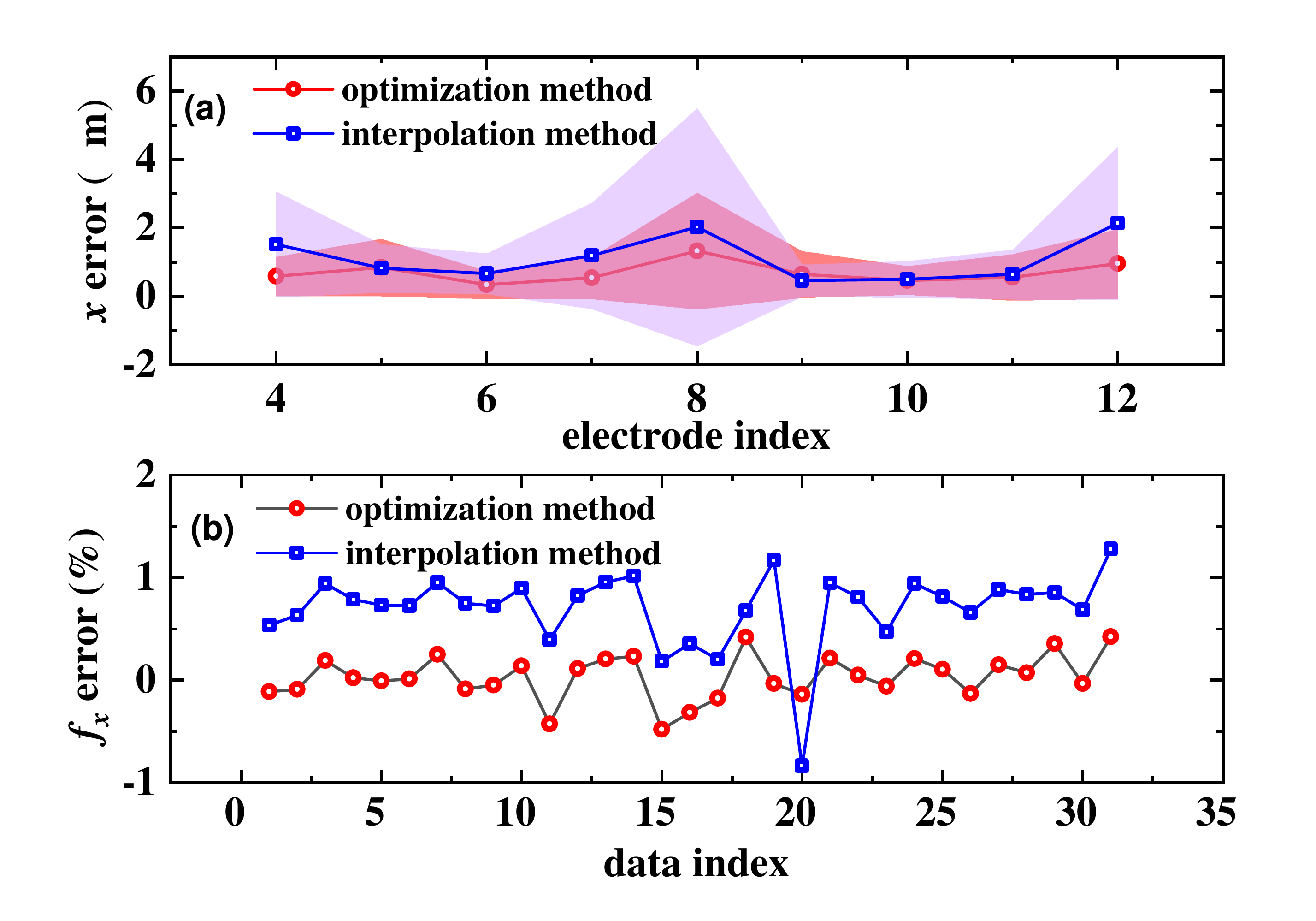}%
	\caption{\label{ErrorOfModels}{ Errors of the $x$ positions and axial secular motion frequencies predicted by the two different trap models. (a) Mean errors of the predicted $x$ positions, with the standard deviation  represented by shades. (b) The relative errors of the axial secular frequencies are plotted. The blue dashed line with diamond (red solid line with dot) is according to the trap model derived using modified interpolation (optimization) method.}}
\end{figure}

Simultaneously, systematic errors induced by height variation, denoted as $\Delta E_x$, are calculated using the 3D simulated potential, as shown in Fig. \ref{RandomErrAndSysErr} with red marks. They are deduced by evaluating the height variation resulting from the adjusted voltages in each measurement. Like the random errors, it is also independent of the data processing method. In our experiments, while this error is not very significant, it is not negligible when compared to random errors. This error may contribute to the positional inaccuracies predicted by the obtained potential model, as shown is Fig. \ref{ErrorOfModels}(a).

It is crucial to emphasize that this error graph specifically describe errors associated with each image data (collected by the adjusted electrode), rather than the error of the unit-voltage potential for each electrode. Due to the comprehensive nature of systematic errors in the calibrated potential, it is impractical to deduce systematic errors of the derived field model. Hence, ensuring a consistent ion trapping height through optimized voltage configuration during measurement is essential. Nevertheless, discussions concerning the ramifications and considerations of this strategy are beyond the scope of this work.

~\\
\textbf{\emph{Assessing the model accuracy }}

The calculated values are subsequently compared with the measured results to assess their accuracy. Employing the trap models established above, one can simulate the equilibrium position of each ion in a linear chain using either the simulated annealing method \cite{wu2017determining} or molecular dynamics simulation \cite{Zhang2007MD}. In our 1D molecular dynamics simulation, we employ the velocity-Verlet algorithm, implementing significant damping to accelerate the equilibrium process. To evaluate the accuracy of the derived trap models, the equilibrium positions of ions in a string and the secular frequencies are determined using the two derived trap models under experimental voltages.

For each electrode $k$, we select five voltage settings from experimental data with different $V_k$ values to conduct molecular dynamics simulations. The chosen $V_k$ values include the maximum and minimum experimental values, with the spacings as evenly distributed as possible. The simulated equilibrium positions are then compared with the measured ones. The position errors of ions belonging to the same voltage-varying electrode are averaged and shown in Fig. \ref{ErrorOfModels}(a), with the shadow representing standard deviation.

The results indicate that the trap model derived by the optimization method is more accurate compared to the one obtained by the modified interpolation, as the errors are generally smaller, along with a reduced standard deviation. The maximum error is approximately 1.2 $\mu$m for the optimization method and 2.2 $\mu$m for the interpolation method.

Relatively larger errors are observed associate with the $8^{th}$ electrode in both trap models. This suggests the presence of common systematic errors in these models. They are possibly stemming from ion height variation during data collection due to the voltage update strategy. Therefore, it is recommended to adopt different voltage configurations in the calibration process rather than solely focusing on variations around a specific set of voltages.

In Fig. \ref{ErrorOfModels}(b), the relative errors of predicted axial secular frequencies calculated using Eq. (\ref{omegax}) for the two different models are displayed. It's worth noting that the secular frequencies of the first 20 points are used to calculate the optimization target $t_2$, and the last 11 points are included for validation purposes. The general trend of the two curves is quite similar, but the errors derived by the modified interpolation method exhibit an offset of about $0.75\%$. It is reasonable to believe that the contribution of the optimization target $t_2$ helps mitigate this offset error, which may be caused by the magnification error in the imaging system.

The trap model derived by the optimization method demonstrates high accuracy in predicting the secular motion frequency, with errors consistently below 0.5$\%$. Its robustness is evident when the trapping conditions extend beyond the experimental region where we derived the trap model. For instance, the secular frequencies used for optimization ranged from 190 to 380 kHz, and when extended to 600 kHz, the error of the predicted frequency remains within 0.5$\%$.

\begin{figure}[thbp]
	\centering
	\includegraphics[width=0.95\linewidth,scale=1.000]{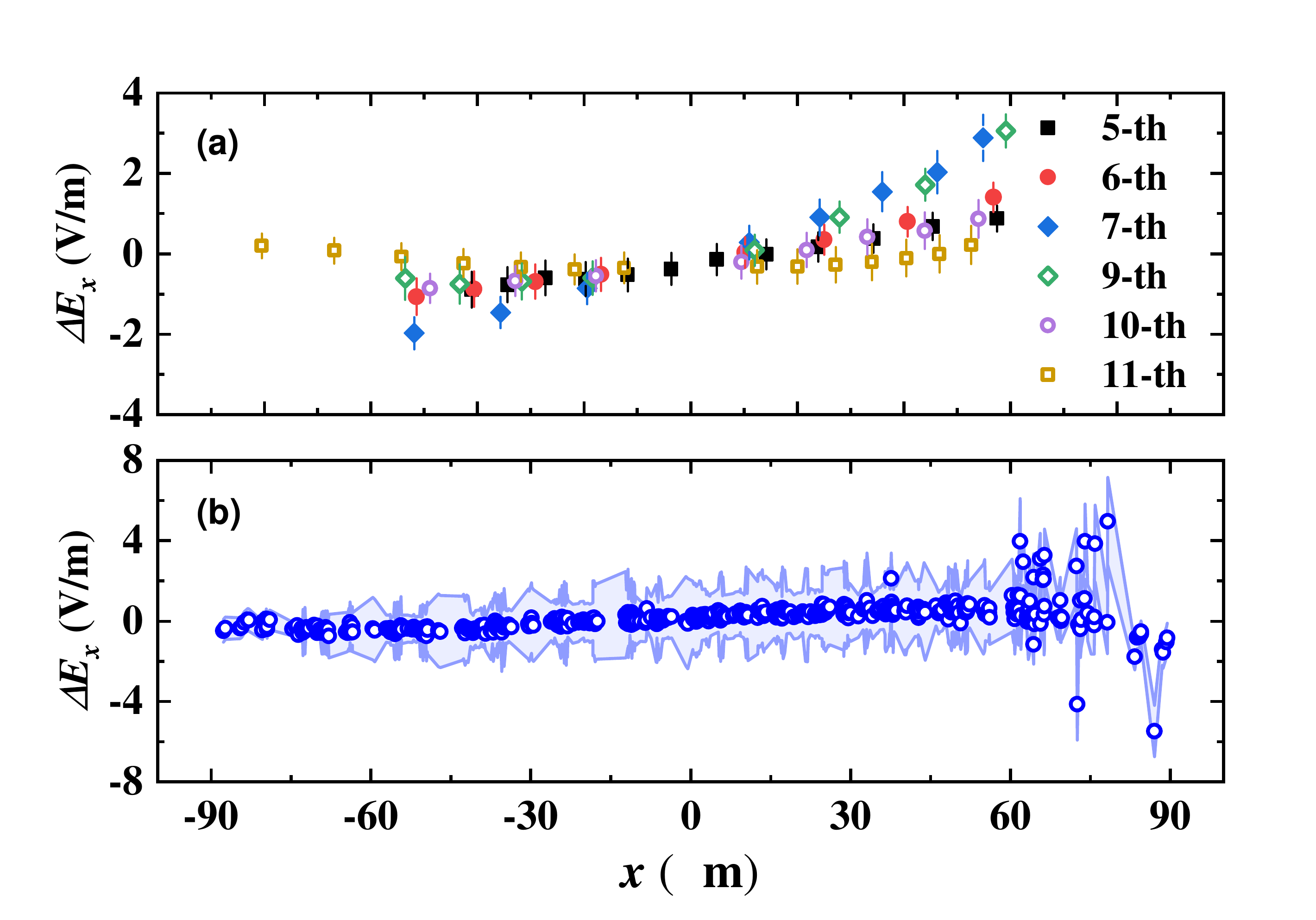}%
	\caption{\label{ModelErr2-xs}{ Results of the reexamination. The uncertainties are attributed to ion position resolution. (a) The difference calculated by single trapped ion under different voltages. (b) The difference calculated by strings of trapped ions under different voltages.}}
\end{figure} 

~\\
\textbf{\emph{A Reexamination experiment}}

To validate the generalization ability of our calibrated potential, we conducted a reexamination by collecting single-ion and ion string images under certain voltages and compared them with the values predicted by the calibrated model, 24 months after the initial calibration date. Over this sufficiently long period, we believe that the stray field undergoes significant variations. The potential is created using the 5th to 11th pairs of electrodes, while the 4th and 12th electrodes are set to $0$ V. Consequently, the voltage settings for the test are very different from those used for calibration and the trapping heights are varied.
To rule out the stray field variations, a compensation function is added to the model to minimize the ion string's position variation from observation. Indeed, this is not the most accurate way and ignored the height variation, but it is the more efficient way. 

For a single trapped ion, the field strength difference can be calculated using the position differences $\delta x$ and the partial derivative of the electric field, expressed as $\delta E_x = \frac{\partial E_x}{\partial x} \delta x$. In the case of ion strings, the electric field is calculated similarly to Eq. (\ref{Fion}). The deviation from the predicted electric field intensity is shown in Fig. \ref{ModelErr2-xs}. It reveals that the error deviation of the electric field calculated by the ion string is smaller than that by a single trapped ion, except for the measurement noise at the ends, which can be understood since the deviation is minimized to recalibrate the stray field. The field deviation of the single-ion probe is relatively larger and electrode-dependent, consistent with the evaluation in Fig. \ref{ErrorOfModels}, as a 1 V/m deviation leads to a 0.3 $\mu$m displacement in a 0.45 MHz harmonic trap. This deviation is related to systematic errors in the model and errors in canceling drifted stray fields. It's important to note that the ion height variation in practical applications may differ from that in the calibration procedure. Nevertheless, in general, the electric field predicted by the derived model aligns well with the measured values, and the model's ability to generalize is acceptable.

\section{Conclusion And Discussion}
\label{sectionV}
A multi-objective optimization method has been introduced for characterizing the potential of a SET. This approach combines the advantages of BEM simulation and experimental measurement and lead to smooth and accurate result.
It naturally accommodates various types of data, such as positions of ions in strings, secular frequencies and positions of single trapped ions under different trapping voltages. Consequently, it mitigates systematic errors of different origins, ensuring higher accuracy in predicting trap frequency and spatial field compared to existing methods. This is verified by comparing the errors of predicted equilibrium positions and the secular frequencies with those derived by the modified interpolation method. The accuracy of the calibrated model is reexamined 24 months later to ensure its precision. It can be further improved, through the adoption of a proper voltage update strategy to maintain the ion height constant during data collection. This strategy, however, need to be studied further to achieve best accuracy.

Our method relies on the parametric expression of electric field intensity. The Lorentz function is found to be accurately enough for rectangular electrode with a width less than 200 $\mu$m. While initially developed for the SET system, we believe it can also be applied to segmented 3D traps, noting that the empirical expression of electric potential needs be replaced. Our method generally requires that the stray field keeps constant during the data acquisition period, and then the 1d stray electric field intensity can be determined. Fortunately, this procedure can be automated to reduce the time taken for data collection and processing, and the total procedure can be completed in an hour after optimization. This may provide a useful tool for calibrating the stray field after the trap potential has been calibrated.  However, when dealing with a large number of electrodes, the experimental period may be prolonged, and the global optimization algorithm may become less efficient. To address this, electrodes can be divided into several groups, and each group can be characterized separately. 

The capability to establish an accurate trap model provides a practical tool for precise trapping potential control. This can find applications in creating evenly spaced linear ion strings, generating trapping wells of special shapes, and precisely controlling the shuttling voltages. These are essential tools in the field of trapped ion quantum simulation and computation.

\begin{acknowledgments}
	
This work is supported by the National Natural Science Foundation of China under Grant Nos. 11904402, 12204543, 12004430, 12074433, 12174447, and 12174448.

\end{acknowledgments}

\section*{Appendix}
\setcounter{figure}{0}
\setcounter{equation}{0}
%定义编号格式，在数字序号前加字符“A"
\renewcommand{\thefigure}{A\arabic{figure}}
\renewcommand{\theequation}{A\arabic{equation}}

The electric potential along the trap axis generated by a single rectangular electrode can be fitted using an appropriate single-peaked function. The quality of the fit depends on the region of interest where the fitted data located. Taking the 8-th pair of electrodes in our SET as an example, it consists of a pair of rectangular electrodes with a separation of 514 $\mu$m, each with a width of 147 $\mu$m and a length of 0.94 mm. The potential along trap axis can be fitted using a Lorentzian function, within different region of interest. 

\begin{figure}[thbp]
	\centering
	\includegraphics[width=0.95\linewidth,scale=1.000]{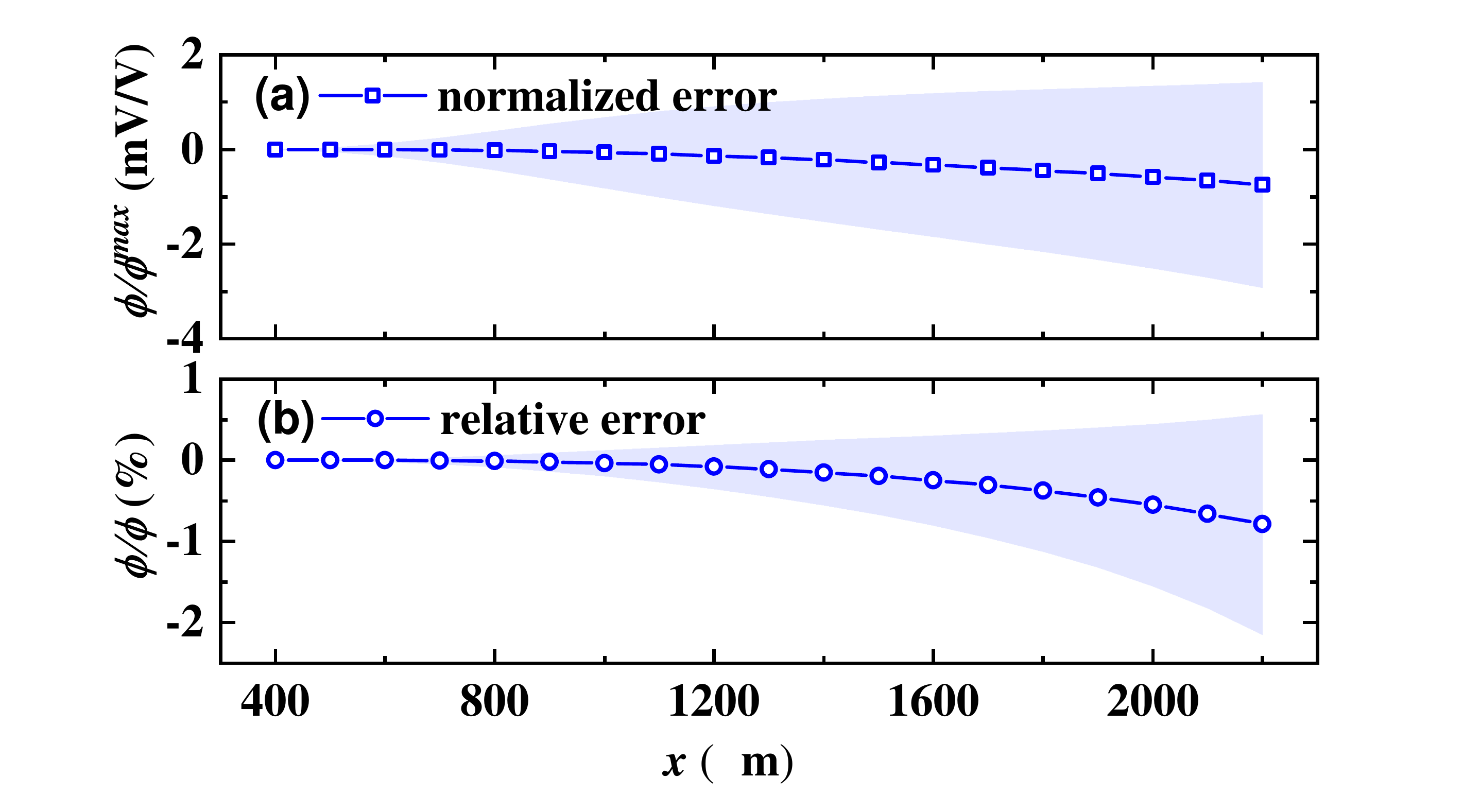}%
	\caption{\label{FitErrVsROI}{ The fitting quality with respect to the region of interest being fitted. (a) The normalized error and (b) the relative error.}}
\end{figure}

To evaluate the fitting qualities, we calculate the difference between the fitted function and the potential at intervals of 1 $\mu$m within the fitting range. The differences are presented in two forms. One is called the normalized error, which is the mean difference divided by the maximum potential according to the fitting range between 400 and 2200 $\mu$m, as shown in Fig. \ref{FitErrVsROI} (a), with its standard deviation depicted in the shaded area. The other is the mean relative error of the fitted model, as shown in Fig. \ref{FitErrVsROI} (b), with its standard deviation also displayed in the shaded area. It can be observed that the average relative errors for fits within 1.4 mm are all within 0.5$\%$, and the normalized errors are less than 1.5 mV/V.

We found the Lorentzian function is primarily suitable for narrow electrodes. For wider ones, a combination of Lorentzian and Gaussian functions will achieve higher fitting precision, with the form:
\begin{equation}
	\phi_k (x) = \frac{a_k \gamma_k }{\gamma_k ^2+x^2}+b_k \exp \left(-\frac{x^2}{\omega_k }\right).
\end{equation} 

To demonstrate this, we utilize the analytical method to calculate the potential of the electrodes. The spatial distribution of the potential calculated in this way is very close to the results of BEM simulations, with the main differences being peak height and width. Using analytical expressions to assess the fitting quality is considered reliable.

\begin{figure}[thbp]
	\centering
	\includegraphics[width=0.95\linewidth,scale=1.000]{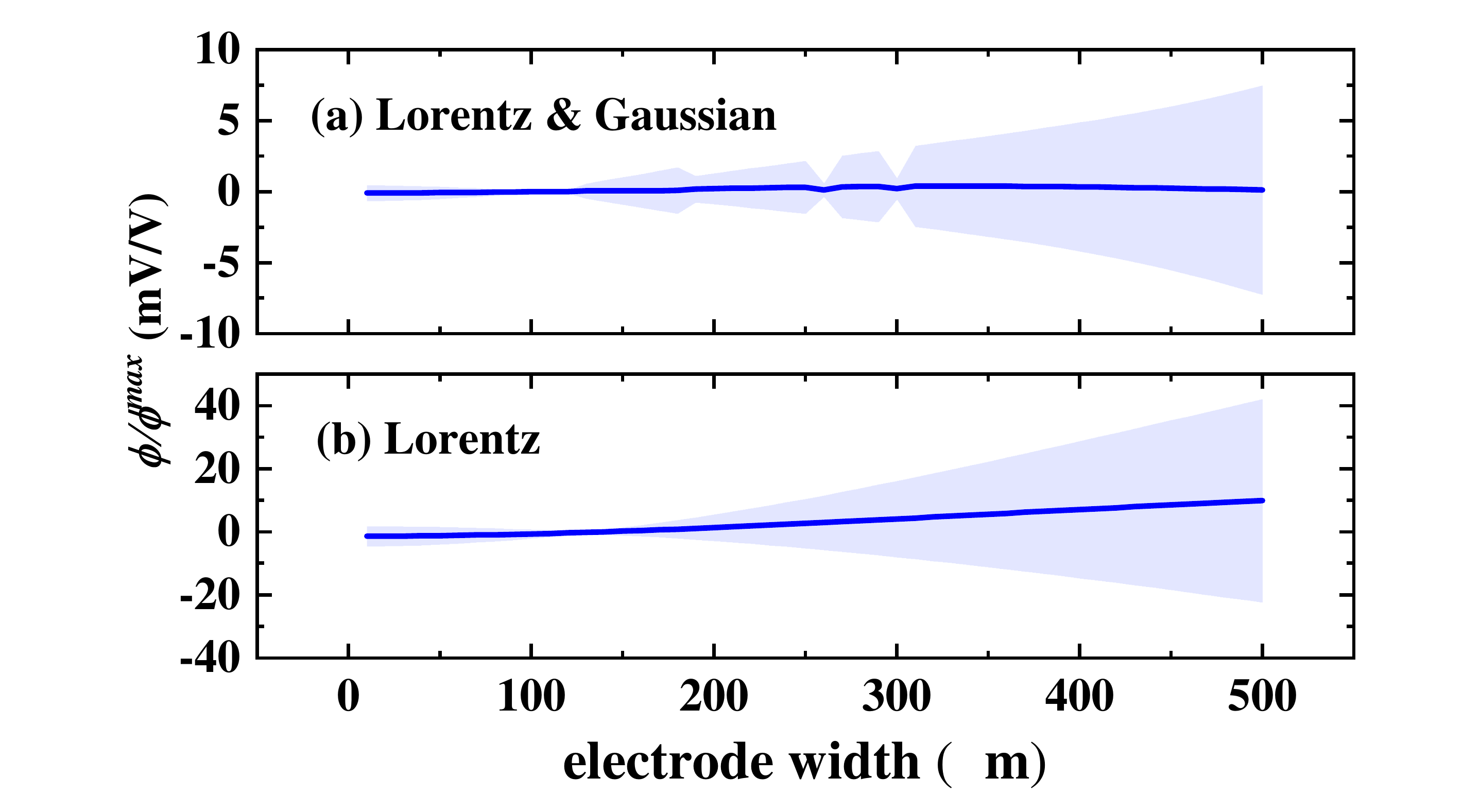}%
	\caption{\label{ErrVsWid}{ The fitting quality with respect to electrode width. (a) The combination of Lorentz function and Gaussian function can provide more fitting degrees of freedom and achieve better results. (b) Lorentz function is well-matched for electrodes less than 200 $\mu$m.}}
\end{figure} 

Considering an pair of electrode each with a length of 3 mm, a spacing of 200 micrometers, and an ion height of 150 micrometers. We gradually vary the electrode width from 10 $\mu$m to 500 $\mu$m and use the data in the middle 2000 $\mu$m for fitting. We then calculate the normalized error for different fitting model, as is shown in Fig. \ref{ErrVsWid}. It is observed that for the electrodes with width less than 200 $\mu$m, the Lorentzian function already provides satisfactory results. As SETs are becoming increasingly miniaturized, electrode widths are typically less than 200 micrometers nowadays. In our trap, the electrode width is 147 micrometers, which falls within the range where Lorentzian function fitting errors are relatively small.
Nevertheless, for electrodes with width less than 500 $\mu$m our method is still applicable with the addition of Gaussian functions.

% Create the reference section using BibTeX:
\bibliographystyle{apsrev4-1}
\bibliography{ModelingSETRef}

%merlin.mbs apsrev4-1.bst 2010-07-25 4.21a (PWD, AO, DPC) hacked
%Control: key (0)
%Control: author (72) initials jnrlst
%Control: editor formatted (1) identically to author
%Control: production of article title (-1) disabled
%Control: page (0) single
%Control: year (1) truncated
%Control: production of eprint (0) enabled
\begin{thebibliography}{36}%
\makeatletter
\providecommand \@ifxundefined [1]{%
 \@ifx{#1\undefined}
}%
\providecommand \@ifnum [1]{%
 \ifnum #1\expandafter \@firstoftwo
 \else \expandafter \@secondoftwo
 \fi
}%
\providecommand \@ifx [1]{%
 \ifx #1\expandafter \@firstoftwo
 \else \expandafter \@secondoftwo
 \fi
}%
\providecommand \natexlab [1]{#1}%
\providecommand \enquote  [1]{``#1''}%
\providecommand \bibnamefont  [1]{#1}%
\providecommand \bibfnamefont [1]{#1}%
\providecommand \citenamefont [1]{#1}%
\providecommand \href@noop [0]{\@secondoftwo}%
\providecommand \href [0]{\begingroup \@sanitize@url \@href}%
\providecommand \@href[1]{\@@startlink{#1}\@@href}%
\providecommand \@@href[1]{\endgroup#1\@@endlink}%
\providecommand \@sanitize@url [0]{\catcode `\\12\catcode `\$12\catcode
  `\&12\catcode `\#12\catcode `\^12\catcode `\_12\catcode `\%12\relax}%
\providecommand \@@startlink[1]{}%
\providecommand \@@endlink[0]{}%
\providecommand \url  [0]{\begingroup\@sanitize@url \@url }%
\providecommand \@url [1]{\endgroup\@href {#1}{\urlprefix }}%
\providecommand \urlprefix  [0]{URL }%
\providecommand \Eprint [0]{\href }%
\providecommand \doibase [0]{http://dx.doi.org/}%
\providecommand \selectlanguage [0]{\@gobble}%
\providecommand \bibinfo  [0]{\@secondoftwo}%
\providecommand \bibfield  [0]{\@secondoftwo}%
\providecommand \translation [1]{[#1]}%
\providecommand \BibitemOpen [0]{}%
\providecommand \bibitemStop [0]{}%
\providecommand \bibitemNoStop [0]{.\EOS\space}%
\providecommand \EOS [0]{\spacefactor3000\relax}%
\providecommand \BibitemShut  [1]{\csname bibitem#1\endcsname}%
\let\auto@bib@innerbib\@empty
%</preamble>
\bibitem [{\citenamefont {Wang}\ \emph {et~al.}(2017)\citenamefont {Wang},
  \citenamefont {Um}, \citenamefont {Zhang}, \citenamefont {An}, \citenamefont
  {Lyu}, \citenamefont {Zhang}, \citenamefont {Duan}, \citenamefont {Yum},\
  and\ \citenamefont {Kim}}]{wang2017ten-minute}%
  \BibitemOpen
  \bibfield  {author} {\bibinfo {author} {\bibfnamefont {Y.}~\bibnamefont
  {Wang}}, \bibinfo {author} {\bibfnamefont {M.}~\bibnamefont {Um}}, \bibinfo
  {author} {\bibfnamefont {J.}~\bibnamefont {Zhang}}, \bibinfo {author}
  {\bibfnamefont {S.}~\bibnamefont {An}}, \bibinfo {author} {\bibfnamefont
  {M.}~\bibnamefont {Lyu}}, \bibinfo {author} {\bibfnamefont {J.-N.}\
  \bibnamefont {Zhang}}, \bibinfo {author} {\bibfnamefont {L.-M.}\ \bibnamefont
  {Duan}}, \bibinfo {author} {\bibfnamefont {D.}~\bibnamefont {Yum}}, \ and\
  \bibinfo {author} {\bibfnamefont {K.}~\bibnamefont {Kim}},\ }\href {\doibase
  10.1038/s41566-017-0007-1} {\bibfield  {journal} {\bibinfo  {journal} {Nature
  Photonics}\ }\textbf {\bibinfo {volume} {11}},\ \bibinfo {pages} {646}
  (\bibinfo {year} {2017})}\BibitemShut {NoStop}%
\bibitem [{\citenamefont {Wang}\ \emph {et~al.}(2021)\citenamefont {Wang},
  \citenamefont {Luan}, \citenamefont {Qiao}, \citenamefont {Um}, \citenamefont
  {Zhang}, \citenamefont {Wang}, \citenamefont {Yuan}, \citenamefont {Gu},
  \citenamefont {Zhang},\ and\ \citenamefont {Kim}}]{wang2021one-hour}%
  \BibitemOpen
  \bibfield  {author} {\bibinfo {author} {\bibfnamefont {P.}~\bibnamefont
  {Wang}}, \bibinfo {author} {\bibfnamefont {C.-Y.}\ \bibnamefont {Luan}},
  \bibinfo {author} {\bibfnamefont {M.}~\bibnamefont {Qiao}}, \bibinfo {author}
  {\bibfnamefont {M.}~\bibnamefont {Um}}, \bibinfo {author} {\bibfnamefont
  {J.}~\bibnamefont {Zhang}}, \bibinfo {author} {\bibfnamefont
  {Y.}~\bibnamefont {Wang}}, \bibinfo {author} {\bibfnamefont {X.}~\bibnamefont
  {Yuan}}, \bibinfo {author} {\bibfnamefont {M.}~\bibnamefont {Gu}}, \bibinfo
  {author} {\bibfnamefont {J.}~\bibnamefont {Zhang}}, \ and\ \bibinfo {author}
  {\bibfnamefont {K.}~\bibnamefont {Kim}},\ }\href {\doibase
  10.1038/s41467-020-20330-w} {\bibfield  {journal} {\bibinfo  {journal}
  {Nature communications}\ }\textbf {\bibinfo {volume} {12}},\ \bibinfo {pages}
  {1} (\bibinfo {year} {2021})}\BibitemShut {NoStop}%
\bibitem [{\citenamefont {Ballance}\ \emph {et~al.}(2016)\citenamefont
  {Ballance}, \citenamefont {Harty}, \citenamefont {Linke}, \citenamefont
  {Sepiol},\ and\ \citenamefont {Lucas}}]{Ballance2016QG-Ca}%
  \BibitemOpen
  \bibfield  {author} {\bibinfo {author} {\bibfnamefont {C.~J.}\ \bibnamefont
  {Ballance}}, \bibinfo {author} {\bibfnamefont {T.~P.}\ \bibnamefont {Harty}},
  \bibinfo {author} {\bibfnamefont {N.~M.}\ \bibnamefont {Linke}}, \bibinfo
  {author} {\bibfnamefont {M.~A.}\ \bibnamefont {Sepiol}}, \ and\ \bibinfo
  {author} {\bibfnamefont {D.~M.}\ \bibnamefont {Lucas}},\ }\href {\doibase
  10.1103/PhysRevLett.117.060504} {\bibfield  {journal} {\bibinfo  {journal}
  {Phys. Rev. Lett.}\ }\textbf {\bibinfo {volume} {117}},\ \bibinfo {pages}
  {060504} (\bibinfo {year} {2016})}\BibitemShut {NoStop}%
\bibitem [{\citenamefont {Gaebler}\ \emph {et~al.}(2016)\citenamefont
  {Gaebler}, \citenamefont {Tan}, \citenamefont {Lin}, \citenamefont {Wan},
  \citenamefont {Bowler}, \citenamefont {Keith}, \citenamefont {Glancy},
  \citenamefont {Coakley}, \citenamefont {Knill}, \citenamefont {Leibfried},\
  and\ \citenamefont {Wineland}}]{Gaebler2016QG-Be}%
  \BibitemOpen
  \bibfield  {author} {\bibinfo {author} {\bibfnamefont {J.~P.}\ \bibnamefont
  {Gaebler}}, \bibinfo {author} {\bibfnamefont {T.~R.}\ \bibnamefont {Tan}},
  \bibinfo {author} {\bibfnamefont {Y.}~\bibnamefont {Lin}}, \bibinfo {author}
  {\bibfnamefont {Y.}~\bibnamefont {Wan}}, \bibinfo {author} {\bibfnamefont
  {R.}~\bibnamefont {Bowler}}, \bibinfo {author} {\bibfnamefont {A.~C.}\
  \bibnamefont {Keith}}, \bibinfo {author} {\bibfnamefont {S.}~\bibnamefont
  {Glancy}}, \bibinfo {author} {\bibfnamefont {K.}~\bibnamefont {Coakley}},
  \bibinfo {author} {\bibfnamefont {E.}~\bibnamefont {Knill}}, \bibinfo
  {author} {\bibfnamefont {D.}~\bibnamefont {Leibfried}}, \ and\ \bibinfo
  {author} {\bibfnamefont {D.~J.}\ \bibnamefont {Wineland}},\ }\href {\doibase
  10.1103/PhysRevLett.117.060505} {\bibfield  {journal} {\bibinfo  {journal}
  {Phys. Rev. Lett.}\ }\textbf {\bibinfo {volume} {117}},\ \bibinfo {pages}
  {060505} (\bibinfo {year} {2016})}\BibitemShut {NoStop}%
\bibitem [{\citenamefont {Srinivas}\ \emph {et~al.}(2021)\citenamefont
  {Srinivas}, \citenamefont {Burd}, \citenamefont {Knaack}, \citenamefont
  {Sutherland}, \citenamefont {Kwiatkowski}, \citenamefont {Glancy},
  \citenamefont {Knill}, \citenamefont {Wineland}, \citenamefont {Leibfried},
  \citenamefont {Wilson} \emph {et~al.}}]{srinivas2021Laser-free}%
  \BibitemOpen
  \bibfield  {author} {\bibinfo {author} {\bibfnamefont {R.}~\bibnamefont
  {Srinivas}}, \bibinfo {author} {\bibfnamefont {S.}~\bibnamefont {Burd}},
  \bibinfo {author} {\bibfnamefont {H.}~\bibnamefont {Knaack}}, \bibinfo
  {author} {\bibfnamefont {R.}~\bibnamefont {Sutherland}}, \bibinfo {author}
  {\bibfnamefont {A.}~\bibnamefont {Kwiatkowski}}, \bibinfo {author}
  {\bibfnamefont {S.}~\bibnamefont {Glancy}}, \bibinfo {author} {\bibfnamefont
  {E.}~\bibnamefont {Knill}}, \bibinfo {author} {\bibfnamefont
  {D.}~\bibnamefont {Wineland}}, \bibinfo {author} {\bibfnamefont
  {D.}~\bibnamefont {Leibfried}}, \bibinfo {author} {\bibfnamefont {A.~C.}\
  \bibnamefont {Wilson}},  \emph {et~al.},\ }\href {\doibase
  10.1038/s41586-021-03809-4} {\bibfield  {journal} {\bibinfo  {journal}
  {Nature}\ }\textbf {\bibinfo {volume} {597}},\ \bibinfo {pages} {209}
  (\bibinfo {year} {2021})}\BibitemShut {NoStop}%
\bibitem [{\citenamefont {Linke}\ \emph {et~al.}(2017)\citenamefont {Linke},
  \citenamefont {Maslov}, \citenamefont {Roetteler}, \citenamefont {Debnath},
  \citenamefont {Figgatt}, \citenamefont {Landsman}, \citenamefont {Wright},\
  and\ \citenamefont {Monroe}}]{linke2017comparison}%
  \BibitemOpen
  \bibfield  {author} {\bibinfo {author} {\bibfnamefont {N.~M.}\ \bibnamefont
  {Linke}}, \bibinfo {author} {\bibfnamefont {D.}~\bibnamefont {Maslov}},
  \bibinfo {author} {\bibfnamefont {M.}~\bibnamefont {Roetteler}}, \bibinfo
  {author} {\bibfnamefont {S.}~\bibnamefont {Debnath}}, \bibinfo {author}
  {\bibfnamefont {C.}~\bibnamefont {Figgatt}}, \bibinfo {author} {\bibfnamefont
  {K.~A.}\ \bibnamefont {Landsman}}, \bibinfo {author} {\bibfnamefont
  {K.}~\bibnamefont {Wright}}, \ and\ \bibinfo {author} {\bibfnamefont
  {C.}~\bibnamefont {Monroe}},\ }\href {\doibase 10.1073/pnas.1618020114}
  {\bibfield  {journal} {\bibinfo  {journal} {Proceedings of the National
  Academy of Sciences}\ }\textbf {\bibinfo {volume} {114}},\ \bibinfo {pages}
  {3305} (\bibinfo {year} {2017})}\BibitemShut {NoStop}%
\bibitem [{\citenamefont {Lin}\ \emph {et~al.}(2009)\citenamefont {Lin},
  \citenamefont {Zhu}, \citenamefont {Islam}, \citenamefont {Kim},
  \citenamefont {Chang}, \citenamefont {Korenblit}, \citenamefont {Monroe},\
  and\ \citenamefont {Duan}}]{Lin2009}%
  \BibitemOpen
  \bibfield  {author} {\bibinfo {author} {\bibfnamefont {G.-D.}\ \bibnamefont
  {Lin}}, \bibinfo {author} {\bibfnamefont {S.-L.}\ \bibnamefont {Zhu}},
  \bibinfo {author} {\bibfnamefont {R.}~\bibnamefont {Islam}}, \bibinfo
  {author} {\bibfnamefont {K.}~\bibnamefont {Kim}}, \bibinfo {author}
  {\bibfnamefont {M.-S.}\ \bibnamefont {Chang}}, \bibinfo {author}
  {\bibfnamefont {S.}~\bibnamefont {Korenblit}}, \bibinfo {author}
  {\bibfnamefont {C.}~\bibnamefont {Monroe}}, \ and\ \bibinfo {author}
  {\bibfnamefont {L.-M.}\ \bibnamefont {Duan}},\ }\href {\doibase
  10.1209/0295-5075/86/60004} {\bibfield  {journal} {\bibinfo  {journal}
  {Europhysics Letters}\ }\textbf {\bibinfo {volume} {86}},\ \bibinfo {pages}
  {60004} (\bibinfo {year} {2009})}\BibitemShut {NoStop}%
\bibitem [{\citenamefont {Xie}\ \emph {et~al.}(2017)\citenamefont {Xie},
  \citenamefont {Zhang}, \citenamefont {Ou}, \citenamefont {Chen},
  \citenamefont {Zhang}, \citenamefont {Wu}, \citenamefont {Wu},\ and\
  \citenamefont {Chen}}]{Xie2017}%
  \BibitemOpen
  \bibfield  {author} {\bibinfo {author} {\bibfnamefont {Y.}~\bibnamefont
  {Xie}}, \bibinfo {author} {\bibfnamefont {X.}~\bibnamefont {Zhang}}, \bibinfo
  {author} {\bibfnamefont {B.}~\bibnamefont {Ou}}, \bibinfo {author}
  {\bibfnamefont {T.}~\bibnamefont {Chen}}, \bibinfo {author} {\bibfnamefont
  {J.}~\bibnamefont {Zhang}}, \bibinfo {author} {\bibfnamefont
  {C.}~\bibnamefont {Wu}}, \bibinfo {author} {\bibfnamefont {W.}~\bibnamefont
  {Wu}}, \ and\ \bibinfo {author} {\bibfnamefont {P.}~\bibnamefont {Chen}},\
  }\href {\doibase 10.1103/PhysRevA.95.032341} {\bibfield  {journal} {\bibinfo
  {journal} {Phys. Rev. A}\ }\textbf {\bibinfo {volume} {95}},\ \bibinfo
  {pages} {032341} (\bibinfo {year} {2017})}\BibitemShut {NoStop}%
\bibitem [{\citenamefont {Kielpinski}\ \emph {et~al.}(2002)\citenamefont
  {Kielpinski}, \citenamefont {Monroe},\ and\ \citenamefont
  {Wineland}}]{kielpinski2002architecture}%
  \BibitemOpen
  \bibfield  {author} {\bibinfo {author} {\bibfnamefont {D.}~\bibnamefont
  {Kielpinski}}, \bibinfo {author} {\bibfnamefont {C.}~\bibnamefont {Monroe}},
  \ and\ \bibinfo {author} {\bibfnamefont {D.}~\bibnamefont {Wineland}},\
  }\href {\doibase 10.1038/nature00784} {\bibfield  {journal} {\bibinfo
  {journal} {Nature}\ }\textbf {\bibinfo {volume} {417}},\ \bibinfo {pages}
  {709} (\bibinfo {year} {2002})}\BibitemShut {NoStop}%
\bibitem [{\citenamefont {Chiaverini}\ \emph {et~al.}(2005)\citenamefont
  {Chiaverini}, \citenamefont {Blakestad}, \citenamefont {Britton},
  \citenamefont {Jost}, \citenamefont {Langer}, \citenamefont {Leibfried},
  \citenamefont {Ozeri},\ and\ \citenamefont {Wineland}}]{Chiaverini2005SET}%
  \BibitemOpen
  \bibfield  {author} {\bibinfo {author} {\bibfnamefont {J.}~\bibnamefont
  {Chiaverini}}, \bibinfo {author} {\bibfnamefont {R.~B.}\ \bibnamefont
  {Blakestad}}, \bibinfo {author} {\bibfnamefont {J.}~\bibnamefont {Britton}},
  \bibinfo {author} {\bibfnamefont {J.~D.}\ \bibnamefont {Jost}}, \bibinfo
  {author} {\bibfnamefont {C.}~\bibnamefont {Langer}}, \bibinfo {author}
  {\bibfnamefont {D.}~\bibnamefont {Leibfried}}, \bibinfo {author}
  {\bibfnamefont {R.}~\bibnamefont {Ozeri}}, \ and\ \bibinfo {author}
  {\bibfnamefont {D.~J.}\ \bibnamefont {Wineland}},\ }\href
  {https://dl.acm.org/doi/abs/10.5555/2011670.2011671} {\bibfield  {journal}
  {\bibinfo  {journal} {Quantum Info. Comput.}\ }\textbf {\bibinfo {volume}
  {5}},\ \bibinfo {pages} {419} (\bibinfo {year} {2005})}\BibitemShut {NoStop}%
\bibitem [{\citenamefont {Seidelin}\ \emph {et~al.}(2006)\citenamefont
  {Seidelin}, \citenamefont {Chiaverini}, \citenamefont {Reichle},
  \citenamefont {Bollinger}, \citenamefont {Leibfried}, \citenamefont
  {Britton}, \citenamefont {Wesenberg}, \citenamefont {Blakestad},
  \citenamefont {Epstein}, \citenamefont {Hume}, \citenamefont {Itano},
  \citenamefont {Jost}, \citenamefont {Langer}, \citenamefont {Ozeri},
  \citenamefont {Shiga},\ and\ \citenamefont {Wineland}}]{Seidelin2006SET}%
  \BibitemOpen
  \bibfield  {author} {\bibinfo {author} {\bibfnamefont {S.}~\bibnamefont
  {Seidelin}}, \bibinfo {author} {\bibfnamefont {J.}~\bibnamefont
  {Chiaverini}}, \bibinfo {author} {\bibfnamefont {R.}~\bibnamefont {Reichle}},
  \bibinfo {author} {\bibfnamefont {J.~J.}\ \bibnamefont {Bollinger}}, \bibinfo
  {author} {\bibfnamefont {D.}~\bibnamefont {Leibfried}}, \bibinfo {author}
  {\bibfnamefont {J.}~\bibnamefont {Britton}}, \bibinfo {author} {\bibfnamefont
  {J.~H.}\ \bibnamefont {Wesenberg}}, \bibinfo {author} {\bibfnamefont {R.~B.}\
  \bibnamefont {Blakestad}}, \bibinfo {author} {\bibfnamefont {R.~J.}\
  \bibnamefont {Epstein}}, \bibinfo {author} {\bibfnamefont {D.~B.}\
  \bibnamefont {Hume}}, \bibinfo {author} {\bibfnamefont {W.~M.}\ \bibnamefont
  {Itano}}, \bibinfo {author} {\bibfnamefont {J.~D.}\ \bibnamefont {Jost}},
  \bibinfo {author} {\bibfnamefont {C.}~\bibnamefont {Langer}}, \bibinfo
  {author} {\bibfnamefont {R.}~\bibnamefont {Ozeri}}, \bibinfo {author}
  {\bibfnamefont {N.}~\bibnamefont {Shiga}}, \ and\ \bibinfo {author}
  {\bibfnamefont {D.~J.}\ \bibnamefont {Wineland}},\ }\href {\doibase
  10.1103/PhysRevLett.96.253003} {\bibfield  {journal} {\bibinfo  {journal}
  {Phys. Rev. Lett.}\ }\textbf {\bibinfo {volume} {96}},\ \bibinfo {pages}
  {253003} (\bibinfo {year} {2006})}\BibitemShut {NoStop}%
\bibitem [{\citenamefont {Home}\ \emph {et~al.}(2009)\citenamefont {Home},
  \citenamefont {Hanneke}, \citenamefont {Jost}, \citenamefont {Amini},
  \citenamefont {Leibfried},\ and\ \citenamefont
  {Wineland}}]{Jonathan2009QCCD}%
  \BibitemOpen
  \bibfield  {author} {\bibinfo {author} {\bibfnamefont {J.~P.}\ \bibnamefont
  {Home}}, \bibinfo {author} {\bibfnamefont {D.}~\bibnamefont {Hanneke}},
  \bibinfo {author} {\bibfnamefont {J.~D.}\ \bibnamefont {Jost}}, \bibinfo
  {author} {\bibfnamefont {J.~M.}\ \bibnamefont {Amini}}, \bibinfo {author}
  {\bibfnamefont {D.}~\bibnamefont {Leibfried}}, \ and\ \bibinfo {author}
  {\bibfnamefont {D.~J.}\ \bibnamefont {Wineland}},\ }\href {\doibase
  10.1126/science.1177077} {\bibfield  {journal} {\bibinfo  {journal}
  {Science}\ }\textbf {\bibinfo {volume} {325}},\ \bibinfo {pages} {1227}
  (\bibinfo {year} {2009})}\BibitemShut {NoStop}%
\bibitem [{\citenamefont {Pino}\ \emph {et~al.}(2021)\citenamefont {Pino},
  \citenamefont {Dreiling}, \citenamefont {Figgatt}, \citenamefont {Gaebler},
  \citenamefont {Moses}, \citenamefont {Allman}, \citenamefont {Baldwin},
  \citenamefont {Foss-Feig}, \citenamefont {Hayes}, \citenamefont {Mayer},
  \citenamefont {Ryan-Anderson},\ and\ \citenamefont
  {Neyenhuis}}]{pino2021demonstration}%
  \BibitemOpen
  \bibfield  {author} {\bibinfo {author} {\bibfnamefont {J.~M.}\ \bibnamefont
  {Pino}}, \bibinfo {author} {\bibfnamefont {J.~M.}\ \bibnamefont {Dreiling}},
  \bibinfo {author} {\bibfnamefont {C.}~\bibnamefont {Figgatt}}, \bibinfo
  {author} {\bibfnamefont {J.~P.}\ \bibnamefont {Gaebler}}, \bibinfo {author}
  {\bibfnamefont {S.~A.}\ \bibnamefont {Moses}}, \bibinfo {author}
  {\bibfnamefont {M.}~\bibnamefont {Allman}}, \bibinfo {author} {\bibfnamefont
  {C.}~\bibnamefont {Baldwin}}, \bibinfo {author} {\bibfnamefont
  {M.}~\bibnamefont {Foss-Feig}}, \bibinfo {author} {\bibfnamefont
  {D.}~\bibnamefont {Hayes}}, \bibinfo {author} {\bibfnamefont
  {K.}~\bibnamefont {Mayer}}, \bibinfo {author} {\bibfnamefont
  {C.}~\bibnamefont {Ryan-Anderson}}, \ and\ \bibinfo {author} {\bibfnamefont
  {B.}~\bibnamefont {Neyenhuis}},\ }\href {\doibase 10.1038/s41586-021-03318-4}
  {\bibfield  {journal} {\bibinfo  {journal} {Nature}\ }\textbf {\bibinfo
  {volume} {592}},\ \bibinfo {pages} {209} (\bibinfo {year}
  {2021})}\BibitemShut {NoStop}%
\bibitem [{\citenamefont {F{\"u}rst}\ \emph {et~al.}(2014)\citenamefont
  {F{\"u}rst}, \citenamefont {Goerz}, \citenamefont {Poschinger}, \citenamefont
  {Murphy}, \citenamefont {Montangero}, \citenamefont {Calarco}, \citenamefont
  {Schmidt-Kaler}, \citenamefont {Singer},\ and\ \citenamefont
  {Koch}}]{furst2014controlling}%
  \BibitemOpen
  \bibfield  {author} {\bibinfo {author} {\bibfnamefont {H.}~\bibnamefont
  {F{\"u}rst}}, \bibinfo {author} {\bibfnamefont {M.~H.}\ \bibnamefont
  {Goerz}}, \bibinfo {author} {\bibfnamefont {U.}~\bibnamefont {Poschinger}},
  \bibinfo {author} {\bibfnamefont {M.}~\bibnamefont {Murphy}}, \bibinfo
  {author} {\bibfnamefont {S.}~\bibnamefont {Montangero}}, \bibinfo {author}
  {\bibfnamefont {T.}~\bibnamefont {Calarco}}, \bibinfo {author} {\bibfnamefont
  {F.}~\bibnamefont {Schmidt-Kaler}}, \bibinfo {author} {\bibfnamefont
  {K.}~\bibnamefont {Singer}}, \ and\ \bibinfo {author} {\bibfnamefont {C.~P.}\
  \bibnamefont {Koch}},\ }\href {\doibase 10.1088/1367-2630/16/7/075007}
  {\bibfield  {journal} {\bibinfo  {journal} {New Journal of Physics}\ }\textbf
  {\bibinfo {volume} {16}},\ \bibinfo {pages} {075007} (\bibinfo {year}
  {2014})}\BibitemShut {NoStop}%
\bibitem [{\citenamefont {Walther}\ \emph {et~al.}(2012)\citenamefont
  {Walther}, \citenamefont {Ziesel}, \citenamefont {Ruster}, \citenamefont
  {Dawkins}, \citenamefont {Ott}, \citenamefont {Hettrich}, \citenamefont
  {Singer}, \citenamefont {Schmidt-Kaler},\ and\ \citenamefont
  {Poschinger}}]{Walther2012Transport}%
  \BibitemOpen
  \bibfield  {author} {\bibinfo {author} {\bibfnamefont {A.}~\bibnamefont
  {Walther}}, \bibinfo {author} {\bibfnamefont {F.}~\bibnamefont {Ziesel}},
  \bibinfo {author} {\bibfnamefont {T.}~\bibnamefont {Ruster}}, \bibinfo
  {author} {\bibfnamefont {S.~T.}\ \bibnamefont {Dawkins}}, \bibinfo {author}
  {\bibfnamefont {K.}~\bibnamefont {Ott}}, \bibinfo {author} {\bibfnamefont
  {M.}~\bibnamefont {Hettrich}}, \bibinfo {author} {\bibfnamefont
  {K.}~\bibnamefont {Singer}}, \bibinfo {author} {\bibfnamefont
  {F.}~\bibnamefont {Schmidt-Kaler}}, \ and\ \bibinfo {author} {\bibfnamefont
  {U.}~\bibnamefont {Poschinger}},\ }\href {\doibase
  10.1103/PhysRevLett.109.080501} {\bibfield  {journal} {\bibinfo  {journal}
  {Phys. Rev. Lett.}\ }\textbf {\bibinfo {volume} {109}},\ \bibinfo {pages}
  {080501} (\bibinfo {year} {2012})}\BibitemShut {NoStop}%
\bibitem [{\citenamefont {Bowler}\ \emph {et~al.}(2012)\citenamefont {Bowler},
  \citenamefont {Gaebler}, \citenamefont {Lin}, \citenamefont {Tan},
  \citenamefont {Hanneke}, \citenamefont {Jost}, \citenamefont {Home},
  \citenamefont {Leibfried},\ and\ \citenamefont
  {Wineland}}]{Bowler2012Transport}%
  \BibitemOpen
  \bibfield  {author} {\bibinfo {author} {\bibfnamefont {R.}~\bibnamefont
  {Bowler}}, \bibinfo {author} {\bibfnamefont {J.}~\bibnamefont {Gaebler}},
  \bibinfo {author} {\bibfnamefont {Y.}~\bibnamefont {Lin}}, \bibinfo {author}
  {\bibfnamefont {T.~R.}\ \bibnamefont {Tan}}, \bibinfo {author} {\bibfnamefont
  {D.}~\bibnamefont {Hanneke}}, \bibinfo {author} {\bibfnamefont {J.~D.}\
  \bibnamefont {Jost}}, \bibinfo {author} {\bibfnamefont {J.~P.}\ \bibnamefont
  {Home}}, \bibinfo {author} {\bibfnamefont {D.}~\bibnamefont {Leibfried}}, \
  and\ \bibinfo {author} {\bibfnamefont {D.~J.}\ \bibnamefont {Wineland}},\
  }\href {\doibase 10.1103/PhysRevLett.109.080502} {\bibfield  {journal}
  {\bibinfo  {journal} {Phys. Rev. Lett.}\ }\textbf {\bibinfo {volume} {109}},\
  \bibinfo {pages} {080502} (\bibinfo {year} {2012})}\BibitemShut {NoStop}%
\bibitem [{\citenamefont {Ruzic}\ \emph {et~al.}(2022)\citenamefont {Ruzic},
  \citenamefont {Barrick}, \citenamefont {Hunker}, \citenamefont {Law},
  \citenamefont {McFarland}, \citenamefont {McGuinness}, \citenamefont
  {Parazzoli}, \citenamefont {Sterk}, \citenamefont {Van Der~Wall},\ and\
  \citenamefont {Stick}}]{Ruzic2022}%
  \BibitemOpen
  \bibfield  {author} {\bibinfo {author} {\bibfnamefont {B.~P.}\ \bibnamefont
  {Ruzic}}, \bibinfo {author} {\bibfnamefont {T.~A.}\ \bibnamefont {Barrick}},
  \bibinfo {author} {\bibfnamefont {J.~D.}\ \bibnamefont {Hunker}}, \bibinfo
  {author} {\bibfnamefont {R.~J.}\ \bibnamefont {Law}}, \bibinfo {author}
  {\bibfnamefont {B.~K.}\ \bibnamefont {McFarland}}, \bibinfo {author}
  {\bibfnamefont {H.~J.}\ \bibnamefont {McGuinness}}, \bibinfo {author}
  {\bibfnamefont {L.~P.}\ \bibnamefont {Parazzoli}}, \bibinfo {author}
  {\bibfnamefont {J.~D.}\ \bibnamefont {Sterk}}, \bibinfo {author}
  {\bibfnamefont {J.~W.}\ \bibnamefont {Van Der~Wall}}, \ and\ \bibinfo
  {author} {\bibfnamefont {D.}~\bibnamefont {Stick}},\ }\href {\doibase
  10.1103/PhysRevA.105.052409} {\bibfield  {journal} {\bibinfo  {journal}
  {Phys. Rev. A}\ }\textbf {\bibinfo {volume} {105}},\ \bibinfo {pages}
  {052409} (\bibinfo {year} {2022})}\BibitemShut {NoStop}%
\bibitem [{\citenamefont {Sutherland}\ \emph {et~al.}(2021)\citenamefont
  {Sutherland}, \citenamefont {Burd}, \citenamefont {Slichter}, \citenamefont
  {Libby},\ and\ \citenamefont {Leibfried}}]{Sutherland2021Squeezing}%
  \BibitemOpen
  \bibfield  {author} {\bibinfo {author} {\bibfnamefont {R.~T.}\ \bibnamefont
  {Sutherland}}, \bibinfo {author} {\bibfnamefont {S.~C.}\ \bibnamefont
  {Burd}}, \bibinfo {author} {\bibfnamefont {D.~H.}\ \bibnamefont {Slichter}},
  \bibinfo {author} {\bibfnamefont {S.~B.}\ \bibnamefont {Libby}}, \ and\
  \bibinfo {author} {\bibfnamefont {D.}~\bibnamefont {Leibfried}},\ }\href
  {\doibase 10.1103/PhysRevLett.127.083201} {\bibfield  {journal} {\bibinfo
  {journal} {Phys. Rev. Lett.}\ }\textbf {\bibinfo {volume} {127}},\ \bibinfo
  {pages} {083201} (\bibinfo {year} {2021})}\BibitemShut {NoStop}%
\bibitem [{\citenamefont {Oliveira}\ and\ \citenamefont
  {Miranda}(2001)}]{Oliveira2001BSL}%
  \BibitemOpen
  \bibfield  {author} {\bibinfo {author} {\bibfnamefont {M.~H.}\ \bibnamefont
  {Oliveira}}\ and\ \bibinfo {author} {\bibfnamefont {J.~A.}\ \bibnamefont
  {Miranda}},\ }\href {\doibase 10.1088/0143-0807/22/1/304} {\bibfield
  {journal} {\bibinfo  {journal} {European Journal of Physics}\ }\textbf
  {\bibinfo {volume} {22}},\ \bibinfo {pages} {31} (\bibinfo {year}
  {2001})}\BibitemShut {NoStop}%
\bibitem [{\citenamefont {Wesenberg}(2008)}]{Wesenberg2008Els}%
  \BibitemOpen
  \bibfield  {author} {\bibinfo {author} {\bibfnamefont {J.~H.}\ \bibnamefont
  {Wesenberg}},\ }\href {\doibase 10.1103/PhysRevA.78.063410} {\bibfield
  {journal} {\bibinfo  {journal} {Phys. Rev. A}\ }\textbf {\bibinfo {volume}
  {78}},\ \bibinfo {pages} {063410} (\bibinfo {year} {2008})}\BibitemShut
  {NoStop}%
\bibitem [{\citenamefont {House}(2008)}]{House2008Analytic}%
  \BibitemOpen
  \bibfield  {author} {\bibinfo {author} {\bibfnamefont {M.~G.}\ \bibnamefont
  {House}},\ }\href {\doibase 10.1103/PhysRevA.78.033402} {\bibfield  {journal}
  {\bibinfo  {journal} {Phys. Rev. A}\ }\textbf {\bibinfo {volume} {78}},\
  \bibinfo {pages} {033402} (\bibinfo {year} {2008})}\BibitemShut {NoStop}%
\bibitem [{\citenamefont {Singer}\ \emph {et~al.}(2010)\citenamefont {Singer},
  \citenamefont {Poschinger}, \citenamefont {Murphy}, \citenamefont {Ivanov},
  \citenamefont {Ziesel}, \citenamefont {Calarco},\ and\ \citenamefont
  {Schmidt-Kaler}}]{Singer2010RMP}%
  \BibitemOpen
  \bibfield  {author} {\bibinfo {author} {\bibfnamefont {K.}~\bibnamefont
  {Singer}}, \bibinfo {author} {\bibfnamefont {U.}~\bibnamefont {Poschinger}},
  \bibinfo {author} {\bibfnamefont {M.}~\bibnamefont {Murphy}}, \bibinfo
  {author} {\bibfnamefont {P.}~\bibnamefont {Ivanov}}, \bibinfo {author}
  {\bibfnamefont {F.}~\bibnamefont {Ziesel}}, \bibinfo {author} {\bibfnamefont
  {T.}~\bibnamefont {Calarco}}, \ and\ \bibinfo {author} {\bibfnamefont
  {F.}~\bibnamefont {Schmidt-Kaler}},\ }\href {\doibase
  10.1103/RevModPhys.82.2609} {\bibfield  {journal} {\bibinfo  {journal} {Rev.
  Mod. Phys.}\ }\textbf {\bibinfo {volume} {82}},\ \bibinfo {pages} {2609}
  (\bibinfo {year} {2010})}\BibitemShut {NoStop}%
\bibitem [{\citenamefont {Biercuk}\ \emph {et~al.}(2010)\citenamefont
  {Biercuk}, \citenamefont {Uys}, \citenamefont {Britton}, \citenamefont
  {VanDevender},\ and\ \citenamefont {Bollinger}}]{biercuk2010ultrasensitive}%
  \BibitemOpen
  \bibfield  {author} {\bibinfo {author} {\bibfnamefont {M.~J.}\ \bibnamefont
  {Biercuk}}, \bibinfo {author} {\bibfnamefont {H.}~\bibnamefont {Uys}},
  \bibinfo {author} {\bibfnamefont {J.~W.}\ \bibnamefont {Britton}}, \bibinfo
  {author} {\bibfnamefont {A.~P.}\ \bibnamefont {VanDevender}}, \ and\ \bibinfo
  {author} {\bibfnamefont {J.~J.}\ \bibnamefont {Bollinger}},\ }\href {\doibase
  10.1038/nnano.2010.165} {\bibfield  {journal} {\bibinfo  {journal} {Nature
  nanotechnology}\ }\textbf {\bibinfo {volume} {5}},\ \bibinfo {pages} {646}
  (\bibinfo {year} {2010})}\BibitemShut {NoStop}%
\bibitem [{\citenamefont {Gilmore}\ \emph {et~al.}(2021)\citenamefont
  {Gilmore}, \citenamefont {Affolter}, \citenamefont {Lewis-Swan},
  \citenamefont {Barberena}, \citenamefont {Jordan}, \citenamefont {Rey},\ and\
  \citenamefont {Bollinger}}]{Gilmore2021}%
  \BibitemOpen
  \bibfield  {author} {\bibinfo {author} {\bibfnamefont {K.~A.}\ \bibnamefont
  {Gilmore}}, \bibinfo {author} {\bibfnamefont {M.}~\bibnamefont {Affolter}},
  \bibinfo {author} {\bibfnamefont {R.~J.}\ \bibnamefont {Lewis-Swan}},
  \bibinfo {author} {\bibfnamefont {D.}~\bibnamefont {Barberena}}, \bibinfo
  {author} {\bibfnamefont {E.}~\bibnamefont {Jordan}}, \bibinfo {author}
  {\bibfnamefont {A.~M.}\ \bibnamefont {Rey}}, \ and\ \bibinfo {author}
  {\bibfnamefont {J.~J.}\ \bibnamefont {Bollinger}},\ }\href {\doibase
  10.1126/science.abi5226} {\bibfield  {journal} {\bibinfo  {journal}
  {Science}\ }\textbf {\bibinfo {volume} {373}},\ \bibinfo {pages} {673}
  (\bibinfo {year} {2021})},\ \Eprint
  {http://arxiv.org/abs/https://www.science.org/doi/pdf/10.1126/science.abi5226}
  {https://www.science.org/doi/pdf/10.1126/science.abi5226} \BibitemShut
  {NoStop}%
\bibitem [{\citenamefont {Wei}\ \emph {et~al.}(2023)\citenamefont {Wei},
  \citenamefont {Yuan}, \citenamefont {Chen}, \citenamefont {Cui},
  \citenamefont {Li}, \citenamefont {Dai}, \citenamefont {Zhou},\ and\
  \citenamefont {Feng}}]{Feng2023}%
  \BibitemOpen
  \bibfield  {author} {\bibinfo {author} {\bibfnamefont {Y.-Q.}\ \bibnamefont
  {Wei}}, \bibinfo {author} {\bibfnamefont {Q.}~\bibnamefont {Yuan}}, \bibinfo
  {author} {\bibfnamefont {L.}~\bibnamefont {Chen}}, \bibinfo {author}
  {\bibfnamefont {T.-H.}\ \bibnamefont {Cui}}, \bibinfo {author} {\bibfnamefont
  {J.}~\bibnamefont {Li}}, \bibinfo {author} {\bibfnamefont {S.-Q.}\
  \bibnamefont {Dai}}, \bibinfo {author} {\bibfnamefont {F.}~\bibnamefont
  {Zhou}}, \ and\ \bibinfo {author} {\bibfnamefont {M.}~\bibnamefont {Feng}},\
  }\href {\doibase 10.1103/PhysRevApplied.19.064062} {\bibfield  {journal}
  {\bibinfo  {journal} {Phys. Rev. Appl.}\ }\textbf {\bibinfo {volume} {19}},\
  \bibinfo {pages} {064062} (\bibinfo {year} {2023})}\BibitemShut {NoStop}%
\bibitem [{\citenamefont {Berkeland}\ \emph {et~al.}(1998)\citenamefont
  {Berkeland}, \citenamefont {Miller}, \citenamefont {Bergquist}, \citenamefont
  {Itano},\ and\ \citenamefont {Wineland}}]{berkeland1998minimization}%
  \BibitemOpen
  \bibfield  {author} {\bibinfo {author} {\bibfnamefont {D.}~\bibnamefont
  {Berkeland}}, \bibinfo {author} {\bibfnamefont {J.}~\bibnamefont {Miller}},
  \bibinfo {author} {\bibfnamefont {J.~C.}\ \bibnamefont {Bergquist}}, \bibinfo
  {author} {\bibfnamefont {W.~M.}\ \bibnamefont {Itano}}, \ and\ \bibinfo
  {author} {\bibfnamefont {D.~J.}\ \bibnamefont {Wineland}},\ }\href {\doibase
  10.1063/1.367318} {\bibfield  {journal} {\bibinfo  {journal} {Journal of
  applied physics}\ }\textbf {\bibinfo {volume} {83}},\ \bibinfo {pages} {5025}
  (\bibinfo {year} {1998})}\BibitemShut {NoStop}%
\bibitem [{\citenamefont {Harlander}\ \emph {et~al.}(2010)\citenamefont
  {Harlander}, \citenamefont {Brownnutt}, \citenamefont {H{\"a}nsel},\ and\
  \citenamefont {Blatt}}]{harlander2010trapped}%
  \BibitemOpen
  \bibfield  {author} {\bibinfo {author} {\bibfnamefont {M.}~\bibnamefont
  {Harlander}}, \bibinfo {author} {\bibfnamefont {M.}~\bibnamefont
  {Brownnutt}}, \bibinfo {author} {\bibfnamefont {W.}~\bibnamefont
  {H{\"a}nsel}}, \ and\ \bibinfo {author} {\bibfnamefont {R.}~\bibnamefont
  {Blatt}},\ }\href {\doibase 10.1088/1367-2630/12/9/093035} {\bibfield
  {journal} {\bibinfo  {journal} {New Journal of Physics}\ }\textbf {\bibinfo
  {volume} {12}},\ \bibinfo {pages} {093035} (\bibinfo {year}
  {2010})}\BibitemShut {NoStop}%
\bibitem [{\citenamefont {{Wei}}\ \emph {et~al.}(2022)\citenamefont {{Wei}},
  \citenamefont {{Wang}}, \citenamefont {{Liu}}, \citenamefont {{Cui}},
  \citenamefont {{Chen}}, \citenamefont {{Li}}, \citenamefont {{Dai}},
  \citenamefont {{Zhou}},\ and\ \citenamefont {{Feng}}}]{Feng2022}%
  \BibitemOpen
  \bibfield  {author} {\bibinfo {author} {\bibfnamefont {Y.-Q.}\ \bibnamefont
  {{Wei}}}, \bibinfo {author} {\bibfnamefont {Y.-Z.}\ \bibnamefont {{Wang}}},
  \bibinfo {author} {\bibfnamefont {Z.-C.}\ \bibnamefont {{Liu}}}, \bibinfo
  {author} {\bibfnamefont {T.-H.}\ \bibnamefont {{Cui}}}, \bibinfo {author}
  {\bibfnamefont {L.}~\bibnamefont {{Chen}}}, \bibinfo {author} {\bibfnamefont
  {J.}~\bibnamefont {{Li}}}, \bibinfo {author} {\bibfnamefont {S.-Q.}\
  \bibnamefont {{Dai}}}, \bibinfo {author} {\bibfnamefont {F.}~\bibnamefont
  {{Zhou}}}, \ and\ \bibinfo {author} {\bibfnamefont {M.}~\bibnamefont
  {{Feng}}},\ }\href {\doibase 10.1007/s11433-022-1954-7} {\bibfield  {journal}
  {\bibinfo  {journal} {Science China Physics, Mechanics, and Astronomy}\
  }\textbf {\bibinfo {volume} {65}},\ \bibinfo {eid} {110313} (\bibinfo {year}
  {2022})}\BibitemShut {NoStop}%
\bibitem [{\citenamefont {Brownnutt}\ \emph {et~al.}(2015)\citenamefont
  {Brownnutt}, \citenamefont {Kumph}, \citenamefont {Rabl},\ and\ \citenamefont
  {Blatt}}]{Brownnutt2015RMP}%
  \BibitemOpen
  \bibfield  {author} {\bibinfo {author} {\bibfnamefont {M.}~\bibnamefont
  {Brownnutt}}, \bibinfo {author} {\bibfnamefont {M.}~\bibnamefont {Kumph}},
  \bibinfo {author} {\bibfnamefont {P.}~\bibnamefont {Rabl}}, \ and\ \bibinfo
  {author} {\bibfnamefont {R.}~\bibnamefont {Blatt}},\ }\href {\doibase
  10.1103/RevModPhys.87.1419} {\bibfield  {journal} {\bibinfo  {journal} {Rev.
  Mod. Phys.}\ }\textbf {\bibinfo {volume} {87}},\ \bibinfo {pages} {1419}
  (\bibinfo {year} {2015})}\BibitemShut {NoStop}%
\bibitem [{\citenamefont {Huber}\ \emph {et~al.}(2010)\citenamefont {Huber},
  \citenamefont {Ziesel}, \citenamefont {Poschinger}, \citenamefont {Singer},\
  and\ \citenamefont {Schmidt-Kaler}}]{huber2010trapped}%
  \BibitemOpen
  \bibfield  {author} {\bibinfo {author} {\bibfnamefont {G.}~\bibnamefont
  {Huber}}, \bibinfo {author} {\bibfnamefont {F.}~\bibnamefont {Ziesel}},
  \bibinfo {author} {\bibfnamefont {U.}~\bibnamefont {Poschinger}}, \bibinfo
  {author} {\bibfnamefont {K.}~\bibnamefont {Singer}}, \ and\ \bibinfo {author}
  {\bibfnamefont {F.}~\bibnamefont {Schmidt-Kaler}},\ }\href {\doibase
  10.1007/s00340-010-4148-x} {\bibfield  {journal} {\bibinfo  {journal}
  {Applied Physics B}\ }\textbf {\bibinfo {volume} {100}},\ \bibinfo {pages}
  {725} (\bibinfo {year} {2010})}\BibitemShut {NoStop}%
\bibitem [{\citenamefont {Brownnutt}\ \emph {et~al.}(2012)\citenamefont
  {Brownnutt}, \citenamefont {Harlander}, \citenamefont {H{\"a}nsel},\ and\
  \citenamefont {Blatt}}]{brownnutt2012spatially}%
  \BibitemOpen
  \bibfield  {author} {\bibinfo {author} {\bibfnamefont {M.}~\bibnamefont
  {Brownnutt}}, \bibinfo {author} {\bibfnamefont {M.}~\bibnamefont
  {Harlander}}, \bibinfo {author} {\bibfnamefont {W.}~\bibnamefont
  {H{\"a}nsel}}, \ and\ \bibinfo {author} {\bibfnamefont {R.}~\bibnamefont
  {Blatt}},\ }\href {\doibase 10.1007/s00340-012-5032-7} {\bibfield  {journal}
  {\bibinfo  {journal} {Applied Physics B}\ }\textbf {\bibinfo {volume}
  {107}},\ \bibinfo {pages} {1125} (\bibinfo {year} {2012})}\BibitemShut
  {NoStop}%
\bibitem [{\citenamefont {Zhang}\ \emph {et~al.}(2020)\citenamefont {Zhang},
  \citenamefont {Ou}, \citenamefont {Chen}, \citenamefont {Xie}, \citenamefont
  {Wu},\ and\ \citenamefont {Chen}}]{zhang2020versatile}%
  \BibitemOpen
  \bibfield  {author} {\bibinfo {author} {\bibfnamefont {X.}~\bibnamefont
  {Zhang}}, \bibinfo {author} {\bibfnamefont {B.}~\bibnamefont {Ou}}, \bibinfo
  {author} {\bibfnamefont {T.}~\bibnamefont {Chen}}, \bibinfo {author}
  {\bibfnamefont {Y.}~\bibnamefont {Xie}}, \bibinfo {author} {\bibfnamefont
  {W.}~\bibnamefont {Wu}}, \ and\ \bibinfo {author} {\bibfnamefont
  {P.}~\bibnamefont {Chen}},\ }\href {\doibase 10.1088/1402-4896/ab635b}
  {\bibfield  {journal} {\bibinfo  {journal} {Physica Scripta}\ }\textbf
  {\bibinfo {volume} {95}},\ \bibinfo {pages} {045103} (\bibinfo {year}
  {2020})}\BibitemShut {NoStop}%
\bibitem [{\citenamefont {Ou}\ \emph {et~al.}(2016)\citenamefont {Ou},
  \citenamefont {Zhang}, \citenamefont {Zhang}, \citenamefont {Xie},
  \citenamefont {Chen}, \citenamefont {Wu}, \citenamefont {Wu},\ and\
  \citenamefont {Chen}}]{ou2016optimization}%
  \BibitemOpen
  \bibfield  {author} {\bibinfo {author} {\bibfnamefont {B.}~\bibnamefont
  {Ou}}, \bibinfo {author} {\bibfnamefont {J.}~\bibnamefont {Zhang}}, \bibinfo
  {author} {\bibfnamefont {X.}~\bibnamefont {Zhang}}, \bibinfo {author}
  {\bibfnamefont {Y.}~\bibnamefont {Xie}}, \bibinfo {author} {\bibfnamefont
  {T.}~\bibnamefont {Chen}}, \bibinfo {author} {\bibfnamefont {C.}~\bibnamefont
  {Wu}}, \bibinfo {author} {\bibfnamefont {W.}~\bibnamefont {Wu}}, \ and\
  \bibinfo {author} {\bibfnamefont {P.}~\bibnamefont {Chen}},\ }\href {\doibase
  10.1007/s11433-016-0345-5} {\bibfield  {journal} {\bibinfo  {journal}
  {SCIENCE CHINA Physics, Mechanics \& Astronomy}\ }\textbf {\bibinfo {volume}
  {59}},\ \bibinfo {pages} {1} (\bibinfo {year} {2016})}\BibitemShut {NoStop}%
\bibitem [{\citenamefont {Zhang}\ \emph {et~al.}(2017)\citenamefont {Zhang},
  \citenamefont {Xie}, \citenamefont {Liu}, \citenamefont {Ou}, \citenamefont
  {Wu},\ and\ \citenamefont {Chen}}]{zhang2017realizing}%
  \BibitemOpen
  \bibfield  {author} {\bibinfo {author} {\bibfnamefont {J.}~\bibnamefont
  {Zhang}}, \bibinfo {author} {\bibfnamefont {Y.}~\bibnamefont {Xie}}, \bibinfo
  {author} {\bibfnamefont {P.-f.}\ \bibnamefont {Liu}}, \bibinfo {author}
  {\bibfnamefont {B.-q.}\ \bibnamefont {Ou}}, \bibinfo {author} {\bibfnamefont
  {W.}~\bibnamefont {Wu}}, \ and\ \bibinfo {author} {\bibfnamefont {P.-x.}\
  \bibnamefont {Chen}},\ }\href {\doibase 10.1007/s00340-016-6618-2} {\bibfield
   {journal} {\bibinfo  {journal} {Applied Physics B}\ }\textbf {\bibinfo
  {volume} {123}},\ \bibinfo {pages} {1} (\bibinfo {year} {2017})}\BibitemShut
  {NoStop}%
\bibitem [{\citenamefont {Wu}\ \emph {et~al.}(2017)\citenamefont {Wu},
  \citenamefont {Wu}, \citenamefont {Li}, \citenamefont {Ou}, \citenamefont
  {Xie}, \citenamefont {Wu},\ and\ \citenamefont {Chen}}]{wu2017determining}%
  \BibitemOpen
  \bibfield  {author} {\bibinfo {author} {\bibfnamefont {W.-B.}\ \bibnamefont
  {Wu}}, \bibinfo {author} {\bibfnamefont {C.-W.}\ \bibnamefont {Wu}}, \bibinfo
  {author} {\bibfnamefont {J.}~\bibnamefont {Li}}, \bibinfo {author}
  {\bibfnamefont {B.-Q.}\ \bibnamefont {Ou}}, \bibinfo {author} {\bibfnamefont
  {Y.}~\bibnamefont {Xie}}, \bibinfo {author} {\bibfnamefont {W.}~\bibnamefont
  {Wu}}, \ and\ \bibinfo {author} {\bibfnamefont {P.-X.}\ \bibnamefont
  {Chen}},\ }\href {\doibase 10.1088/1674-1056/26/8/080303} {\bibfield
  {journal} {\bibinfo  {journal} {Chinese Physics B}\ }\textbf {\bibinfo
  {volume} {26}},\ \bibinfo {pages} {080303} (\bibinfo {year}
  {2017})}\BibitemShut {NoStop}%
\bibitem [{\citenamefont {Zhang}\ \emph {et~al.}(2007)\citenamefont {Zhang},
  \citenamefont {Offenberg}, \citenamefont {Roth}, \citenamefont {Wilson},\
  and\ \citenamefont {Schiller}}]{Zhang2007MD}%
  \BibitemOpen
  \bibfield  {author} {\bibinfo {author} {\bibfnamefont {C.~B.}\ \bibnamefont
  {Zhang}}, \bibinfo {author} {\bibfnamefont {D.}~\bibnamefont {Offenberg}},
  \bibinfo {author} {\bibfnamefont {B.}~\bibnamefont {Roth}}, \bibinfo {author}
  {\bibfnamefont {M.~A.}\ \bibnamefont {Wilson}}, \ and\ \bibinfo {author}
  {\bibfnamefont {S.}~\bibnamefont {Schiller}},\ }\href {\doibase
  10.1103/PhysRevA.76.012719} {\bibfield  {journal} {\bibinfo  {journal} {Phys.
  Rev. A}\ }\textbf {\bibinfo {volume} {76}},\ \bibinfo {pages} {012719}
  (\bibinfo {year} {2007})}\BibitemShut {NoStop}%
\end{thebibliography}%

\end{document}